\documentclass[aps,onecolumn,preprintnumbers,amsmath,amssymb]{revtex4}
\usepackage{amsmath,amssymb}
\usepackage{dcolumn}
\usepackage{bm}
\usepackage[dvipdfmx]{graphicx,color}
\usepackage[dvipdfmx]{hyperref}
\hypersetup{
colorlinks=true,
urlcolor=blue,
citecolor=blue,
linkcolor=blue
}

\renewcommand{\figurename}{\textbf{Fig.}}
\renewcommand{\thefigure}{\textbf{\arabic{figure}}}

\begin{document}
\title{Observation of the Mott Insulator to Superfluid Crossover of a Driven-Dissipative Bose-Hubbard System}

\author{Takafumi Tomita$^{1}$}\email{tomita@scphys.kyoto-u.ac.jp}
\author{Shuta Nakajima$^{1}$}
\author{Ippei Danshita$^{2}$}
\author{Yosuke Takasu$^{1}$}
\author{Yoshiro Takahashi$^{1}$}
\affiliation{
{$^{1}$Department of Physics, Kyoto University, Kyoto 606-8502, Japan}
\\
{$^{2}$Yukawa Institute for Theoretical Physics, Kyoto University, Kyoto 606-8502, Japan}
}

\begin{abstract}
Dissipation is ubiquitous in nature and plays a crucial role in quantum systems such as causing decoherence of quantum states. Recently, much attention has been paid to an intriguing possibility of dissipation as an efficient tool for preparation and manipulation of quantum states. Here we report the realization of successful demonstration of a novel role of dissipation in a quantum phase transition using cold atoms. We realize an engineered dissipative Bose-Hubbard system by introducing a controllable strength of two-body inelastic collision via photo-association for ultracold bosons in a three-dimensional optical lattice. In the dynamics subjected to a slow ramp-down of the optical lattice, we find that strong on-site dissipation favors the Mott insulating state: the melting of the Mott insulator is delayed and the growth of the phase coherence is suppressed. The controllability of the dissipation is highlighted by quenching the dissipation, providing a novel method for investigating a quantum many-body state and its non-equilibrium dynamics. 

\end{abstract}
\maketitle


\section{Introduction}
Dissipation -coupling to the environment- plays an essential role in quantum systems. 
On the one hand it causes decoherence of quantum states thus it limits the coherent dynamics. Therefore, protection of the quantum states from the coupling to the environment has been a crucial issue in quantum engineering. On the other hand, the dissipation can be used as an efficient tool for preparation and manipulation of particular quantum states of interest~\cite{Daley-14,Muller-12}. Understanding and controlling non-equilibrium dynamics of correlated quantum many-body systems with dissipation are indeed an imperative issue shared in common among experimental systems in diverse areas of physics, including ultracold gases~\cite{Daley-14,Muller-12,Sieberer-16}, Bose-Einstein condensates placed in optical cavities~\cite{Sieberer-16,Ritsch-13}, trapped ions~\cite{Blatt-12, Bohnet-16}, exciton-polariton BEC~\cite{Sieberer-16,Carusotto-13}, microcavity arrays coupled with superconducting qubits~\cite{Houck-12,Fitzpatrick-17}.

Cold atoms, which attract much attention for investigation of many-body quantum systems owing to high controllability of various parameters, are often regarded as an ideal closed (or isolated) quantum system. However, this controllability allows also for creating open quantum systems by introducing dissipation processes. So far, various kinds of theoretical works on the effect of dissipation have predicted novel quantum states engineered by dissipation due to photon scattering and particle loss~\cite{Diehl-08, Witthaut-08, Verstraete-09, Han-09, DallaTorre-10, Diehl-10, Tomadin-11, LeBoite-13, Bidanovic-14, Stannigel-14, Ashida-16}. Experimentally, a one-body dissipation has been introduced in a controlled manner with several methods. The utility of an electron beam has been demonstrated in Ref.~\cite{Barontini-13,Labouvie-15,Labouvie-16}. With a well-designed photon scattering process, measurement backaction on the many-body state~\cite{Patil-15} and the many-body localization in open quantum systems~\cite{Luschen-17} have been investigated. In the case of three-body loss process, controlling the strength of three-body recombination by Feshbach resonance and realization of a novel metastable many-body state have been demonstrated~\cite{Mark-12}. 

Because the two-body interaction is fundamental and crucial for the emergence of the novel quantum states and many-body physics such as quantum phase transitions, it is important to investigate the effect of two-body dissipation on quantum many-body systems in the way that the strength of dissipation can be widely controlled. While the pioneering works were reported in which the two-body loss process was realized by using intrinsic nature of molecules, such as vibrational quenching~\cite{Syassen-08} and chemical reaction~\cite{Yan-13}, and the lifetime of the molecules was investigated, a systematic study of the effect of two-body dissipation on quantum many-body physics has not been reported.

Here we report an investigation of a Bose-Hubbard system using ultracold atoms in a three-dimensional (3D) optical lattice, in which we introduce engineered dissipation of the two-body particle losses. By exploiting the highly controllable nature of the dissipation that we introduce, we successfully reveal the effect of the dissipation on the quantum phase transition from a Mott insulator to a superfluid state in a systematic manner. In particular, we observe in the ramp-down dynamics across the crossover from the Mott insulator to the superfluid states that the melting of the Mott state is delayed and the growth of the phase coherence is suppressed for the strong dissipation. Note that the type of the dissipation introduced in this work is an on-site one. The highly controllable on-site dissipation allows us to study the quench dynamics as the novel method of the initial state preparation, providing a new way for investigating non-equilibrium quantum dynamics. The success in engineering the controllable dissipation of the Bose-Hubbard system offers new opportunities for exploring novel roles of the dissipation in quantum many-body systems.

\section{Results}
\subsection*{Engineered two-body dissipation}
\begin{figure}[tbp]
\includegraphics[width=15cm,clip]{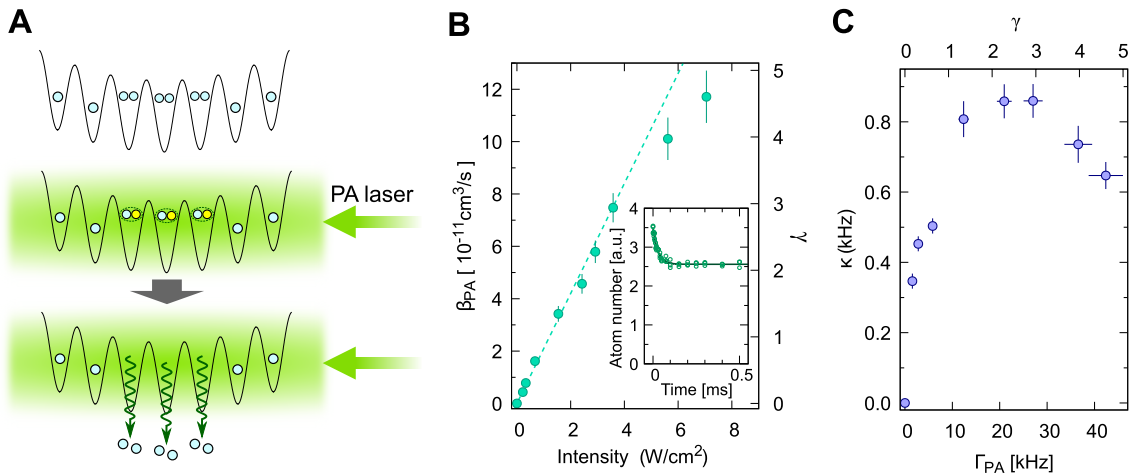}
\caption{
{\bf Engineered dissipation of inelastic two-body collision.} 
\textbf{(A)}, Schematic of the introduced inelastic two-body collision. 
When there are singly- and doubly-occupied sites in the lattice (top), the atoms in the doubly-occupied sites are converted into the molecules by applying the PA laser (middle), and then escape from the lattice due to the high kinetic energy given by the dissociation (bottom). 
\textbf{(B)}, Inelastic collision coefficient $\beta_{\rm{PA}}$ as a function of the intensity of the PA laser. 
The dashed line indicates the linear fit to the low intensity data with a slope of $2.10(7)\times 10^{-11}$ cm$^3$s$^{-1}$/(Wcm$^{-2}$), which well agrees with the theoretical estimation of $2.12 \times 10^{-11}$ cm$^3$s$^{-1}$/(Wcm$^{-2}$)~\cite{Borkowski2010}.
Note that a saturating behavior is observed at the highest intensity, the behavior of which is reported in other experiments performed in a harmonic trap~\cite{0953-4075-39-19-S10,PhysRevA.71.013417,PhysRevA.66.061403,PhysRevLett.101.060406}.
The inset shows time evolution of remaining atom number in a 3D optical lattice for the measurement of the inelastic collision rate $\Gamma_{\rm PA}$. 
The lattice depth is set to $V_0 = 14~E_R$. 
\textbf{(C)}, Inelastic collision rate $\Gamma_{\rm PA}$ dependence of the two-body loss rate $\kappa$ for atoms initially prepared in a Mott insulating state with singly-occupied sites.
The values of $\kappa$ are determined by fitting of the two-body loss function $N(t)=N(0)/(1+\kappa t)$ to the data~\cite{Syassen-08}. 
The scales on the right in \textbf{(B)} and the top in \textbf{(C)} indicate the dimensionless dissipation strength $\gamma$.
}

\label{fig:RateMeasurement}
\end{figure}

A crucial part of the research of a driven-dissipative quantum many-body state is the design of engineered dissipation for a quantum many-body system. In our experiment, two-body inelastic atom loss with controllable strength is successfully implemented by introducing a single-photon photo-association (PA) process for ultracold ytterbium ($^{174}$Yb) atoms in a 3D optical lattice (see Materials and Methods). The PA beam drives the intercombination transition of ${}^1S_0 \leftrightarrow {}^3P_1$, by which two atoms in the doubly-occupied sites are photoassociated into the ${}^1S_0 + {}^3P_1$ molecular state and immediately dissociated into the two ground state atoms (see \textbf{Fig. \ref{fig:RateMeasurement} (A)} and Materials and Methods). This process gives high kinetic energy to the dissociated atoms, which thus results in the escape from the lattice. In this way, the PA laser induces the two-body inelastic collision loss between the two atoms occupying the same site. We can realize controllable strength of inelastic collision coefficient $\beta_{\rm{PA}}$ by changing the intensity of the PA laser $I$, as shown in \textbf{Fig. \ref{fig:RateMeasurement} (B)}. $\beta_{\rm{PA}}$ is determined through the relation $\Gamma_{\rm{PA}}=\beta_{\rm{PA}}\int |w({\bm{r}})|^4 d{\bm{r}}$, where the inelastic collision rate $\Gamma_{\rm{PA}}$ is measured from the exponential fit of collisional loss dynamics as shown in the inset of \textbf{Fig. \ref{fig:RateMeasurement} (B)}, and $w({\bm{r}})$ is the Wannier function of the lowest band. Note that the measurement is done in a deep lattice potential of $V_0 = 14~E_R$, so that atom tunneling is suppressed in the timescale of this measurement. Here $E_R=h^2/(2m\lambda_L^2)$ is a recoil energy with $m$ the mass of the ${}^{174}$Yb atom, $h$ the Planck's constant, and $\lambda_L=532$ nm the wavelength of the lattice laser. In this way, we can realize controllable strength of inelastic collision up to $\beta_{\rm{PA}} \sim 1.2 \times 10^{-10}~{\rm cm}^3/{\rm s} $ corresponding to $\Gamma_{\rm{PA}} \sim $70~kHz in the lattice depth of $V_0 = 14~E_R$ and $\gamma \sim 5$, where $\gamma=\hbar\Gamma_{\rm PA}/U$ is the dimensionless dissipation strength which is independent of the lattice depth and $U$ denotes the on-site interaction.

\subsection*{Model}

It is important to understand how this PA process is effectively described in a dissipative Bose-Hubbard model.
Theoretically, the system of bosonic atoms in a sufficiently deep optical lattice coupled coherently to the ${}^1S_0 + {}^3P_1$ molecular state via the PA laser is well described by the Markovian master equation for the coupled atom-molecule mixture model~\cite{Rousseau-08} with a one-body molecular loss term. By adiabatically eliminating the molecular degrees of freedom on the basis of a second-order perturbation theory for the master equation~\cite{Brennen-05,GarciaRipoll-09}, we derive the effective master equation (see S1 in the Supplementary Materials for details),
\begin{eqnarray}
\hbar \frac{d}{dt}\hat{\rho}_{\rm eff} = -i \left[ 
\hat{H}_{\rm eff}, \hat{\rho}_{\rm eff}
\right]
+ L_2(\hat{\rho}_{\rm eff}),
\label{eq:master_BH_main}
\end{eqnarray}
where
\begin{eqnarray}
\hat{H}_{\rm eff} = \sum_j \frac{U}{2} \hat{n}_{{\rm A},j}(\hat{n}_{{\rm A},j}-1) 
-\sum_{\langle j,k \rangle} J\left( \hat{a}_j^{\dagger} \hat{a}_k + {\rm h.c.} \right),
\end{eqnarray}
\begin{eqnarray}
L_2(\hat{\rho}_{\rm eff}) = \frac{\hbar \Gamma_{\rm PA}}{4}\sum_j \left(
- \hat{a}_j^{\dagger}\hat{a}_j^{\dagger}\hat{a}_j\hat{a}_j \hat{\rho}_{\rm eff}
- \hat{\rho}_{\rm eff} \hat{a}_j^{\dagger}\hat{a}_j^{\dagger}\hat{a}_j\hat{a}_j 
+ 2 \hat{a}_j\hat{a}_j \hat{\rho}_{\rm eff}\hat{a}_j^{\dagger}\hat{a}_j^{\dagger}
\right),
\end{eqnarray}
and 
\begin{eqnarray}
\Gamma_{\rm PA} = 8\frac{g^2}{\hbar^2 \Gamma_{\rm M}}.
\label{eq:geff}
\end{eqnarray}
$\hat{a}_j$ denotes the annihilation operator of atoms at site $j$ and $\hat{n}_{{\rm A},j}=\hat{a}_j^{\dagger}\hat{a}_j$. $\langle j,k\rangle$ represents nearest-neighboring pairs of lattice sites. This model is nothing but the single-component Bose-Hubbard model with a two-body loss term~\cite{GarciaRipoll-09}, where $J$, and $\Gamma_{\rm PA}$ denote the hopping energy, and the strength of the two-body inelastic collision induced by PA. In Eq.~(\ref{eq:geff}), $g$ and $\Gamma_{\rm M}$ denote the strengths of the atom-molecule coupling and the one-body molecular loss, respectively. While $g$ is controllable by varying the intensity of the PA laser, $\Gamma_{\rm M}$ is fixed for a specific molecular state. Note that the effective master equation (\ref{eq:master_BH_main}) is valid only when $\hbar\Gamma_{\rm M} \gg \max(|g|,|D|,|W|,|U|,J)$, where $D$ and $W$ denote the detuning of the PA coupling and the on-site interaction between an atom and a molecule, respectively. Since $\max(|g|,|D|,|W|,|U|,J)/\hbar \sim 100\,{\rm kHz}$ at most and $\Gamma_{\rm M} \sim 1\,{\rm MHz}$ in our experiments (see S2 in the Supplementary Materials), this condition is safely satisfied.

\subsection*{Stability of the atoms with a unit-filling initial state}

\begin{figure}[tbp]
\includegraphics[width=16cm,clip]{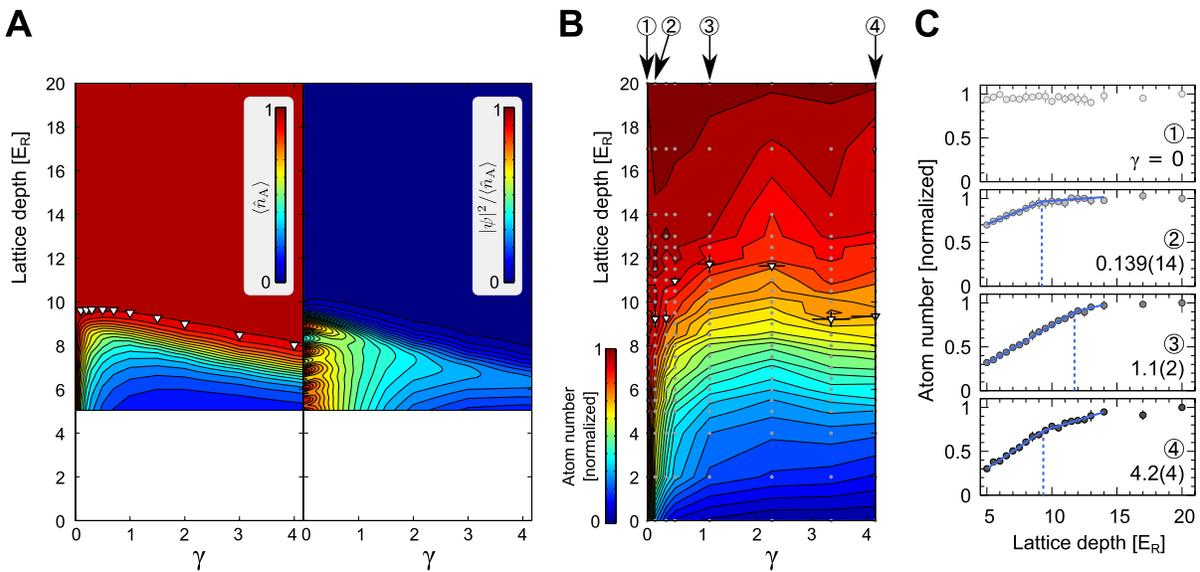}
\caption{
{\bf Atom loss and condensate fraction.} 
\textbf{(A)}, Numerical calculation of the atom number per site $\langle \hat{n}_{\rm A} \rangle$ (left) and the condensate fraction $|\psi|^2/\langle \hat{n}_{\rm A} \rangle$ (right) based on the dissipative Bose-Hubbard model with the Gutzwiller approximation.
The time sequence of the lattice depth and the strength of the dissipation are set to be almost identical to those in the experiments shown in \textbf{(B)} (see also Fig.~S11 of the Supplementary Materials). 
\textbf{(B)}, Atom number diagram. 
The experimental data of the atom number are shown as the gray dots as a function of the final lattice depth for various strengths of dissipation, and are interpolated. 
The white triangles show the lattice depths at which the atom loss sets in, determined from the analysis in \textbf{(C)}. 
The numbers \textcircled{\footnotesize 1} to \textcircled{\footnotesize 4} correspond to the dissipation strengths for which the atom number changes are plotted in \textbf{(C)}. 
\textbf{(C)}, Temporal change of the atom number during a ramp-down sequence for four representative strengths of the dissipation. 
The atom number is normalized by the initial atom number at the lattice depth of $V_0 =20~E_R$. 
Blue lines are double linear fits to extract the onset of the atom loss, which are shown as dotted lines. 
}
\label{fig:AtomLoss}
\end{figure}

Before studying the effect of the dissipation on the quantum phase transition, we investigate the stability of the atoms with a unit-filling initial state at a fixed lattice depth. Here, the strength of the dissipation is varied in a wide range from the weak region, in which the dissipation acts as perturbation, to the strong region, in which it exceeds any other energy scale. In contrast to the previous works in which the experiments were done only in the limited range of the dissipation strength, the wide range of our engineered dissipation enables us to observe a crossover between qualitatively different roles of the dissipation. This gives a clue to understand the many-body physics with the dissipation.

In contrast to the measurement of $\beta_{\rm{PA}}$ shown in \textbf{Fig. \ref{fig:RateMeasurement} (B)}, this measurement  is done at  a shallow lattice depth of $V_0= 8~E_R$ in which the tunneling rate $6J/\hbar$ is 4.7 kHz.
Thus, in the absence of dissipation, atoms can tunnel to neighboring sites in a timescale of this measurement, so that 
we can investigate how the atom tunneling, which is the only mechanism causing doubly-occupied sites, is modified by the dissipation.
In our experiment, we first adiabatically load a Bose-Einstein condensate (BEC) of $ 1.0 \times 10^4$ atoms into a 3D optical lattice with $V_0 = 15~E_R$, in which the state is a singly-occupied Mott insulator. Subsequently, we ramp down the lattice to $V_0 =8~E_R$ in 0.2 ms, and apply the PA laser at the same time.
The initial atom number at each site is at most unity, confirmed by the absence of the atom loss by the PA laser and also the occupancy-sensitive high-resolution laser spectroscopy \cite{Kato-16}.

The measured two-body loss rate $\kappa$ is shown in \textbf{Fig. \ref{fig:RateMeasurement} (C)}.
For weak dissipation, the two-body loss rate $\kappa$ grows as $\Gamma_{\rm{PA}}$ increases, which reflects the increase of the detection rate of tunneling.
For strong dissipation, however, the two-body loss rate decreases when $\Gamma_{\rm{PA}}$ increases. Namely, atom loss is suppressed by the strong on-site dissipation.
This counter-intuitive behavior is a manifestation of the continuous quantum Zeno effect~\cite{misra1977zeno}, that is, the strong two-body inelastic collision plays a role of the strong measurement and suppresses the coherent process of tunneling.
From  the comparison between the theory and experiment in a wide range of dissipation strength, we confirm that the measured loss behavior
 correctly captures the theoretical prediction (see S3 in the Supplementary Materials).
Note that we observe unexpectedly large atom loss for much higher intensity of PA laser, which prevents the suppression of the two-body loss rate for the strong dissipation region over $\gamma \sim 5$ from clear observation (see S4 in the Supplementary Materials). Therefore, in our experiment, we restrict the region of the dissipation strength under $\gamma \sim 5$.

\subsection*{Effect of the dissipation on the quantum phase transition}

\begin{figure}[tbp]
\includegraphics[width=16cm,clip]{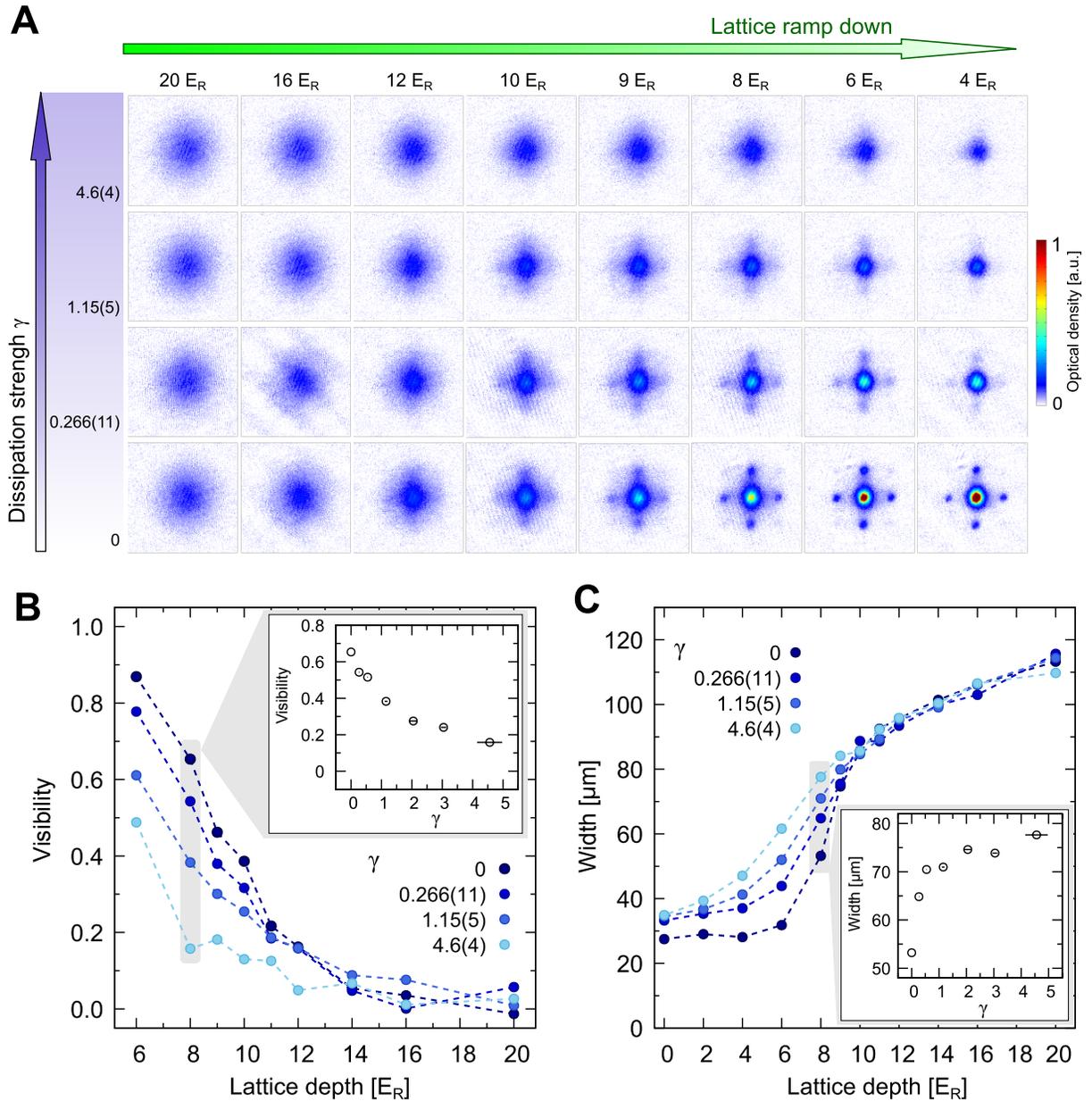}
\caption{
{\bf Coherence properties across the Mott insulator to superfluid crossover.} 
\textbf{(A)}, TOF absorption image. 
The images are taken with different final lattice depths and strengths of the dissipation, and averaged over 20 shots at each parameter. 
\textbf{(B)}, Visibility of the interference peak of the images. 
\textbf{(C)}, Width of the density distribution. 
The width is the FWHM obtained by the Gaussian fitting. The insets in the figures \textbf{(B)} and \textbf{(C)} show the values varying the dissipation strength in the fixed lattice depth of 8 $E_R$.
}
\label{fig:AbsImage}
\end{figure}

We next investigate the effect of the on-site dissipation on the quantum phase transition from the Mott insulator to the superfluid~\cite{greiner2002quantum}, which is the main topic of the present work. 
Specifically, starting with a singly-occupied Mott insulating state, we analyze the dynamics of the atoms subjected to the PA laser during a ramp-down of the lattice depth. 
The ramp-down speed is $-2\,E_{\rm R}/{\rm ms}$, which is much slower than the case of the two-body loss measurement discussed above. 
Before presenting the experimental observation, we theoretically analyze such dynamics by assuming a homogeneous system and solving the effective master equation (\ref{eq:master_BH_main}) within the Gutzwiller mean-field approximation~\cite{Diehl-10,Tomadin-11} in order to obtain some insights on the problem. 
Details of the theoretical analyses are shown in S3 in the Supplementary Materials.

An important effect of the dissipation on the quantum phase transition is that it explicitly breaks the conservation of the particle number of the system. Since the superfluid-Mott insulator transition at $\gamma = 0$ is originated from the U(1) symmetry associated with the particle-number conservation, the introduction of finite $\gamma$ changes the transition to a crossover. Notice, however, that the two-body loss term does not explicitly break the U(1) symmetry. Indeed, the master equation (\ref{eq:master_BH_main}) is invariant under the U(1) transformation, $\hat{a}_j \rightarrow \hat{a}_j e^{i\varphi}$, where $\varphi$ is an arbitrary constant.

This crossover can be theoretically characterized by the growth rate of the superfluid order-parameter amplitude; when the growth rate is smaller, the system is deeper in the Mott insulator region. According to this characterization, we find that in the strong on-site dissipation region, where $\gamma \gg 1$, the Mott insulating state is more favored for larger $\gamma$ (see Fig. S10 (A) of the Supplementary Materials). This effect originates from the quantum Zeno suppression of the tunneling, which is observed in the two-body loss rate measurement.

This interesting effect of the on-site dissipation on the crossover manifests as the delay in the melting of the singly-occupied Mott insulator in the ramp-down dynamics. In \textbf{Figs.~\ref{fig:AtomLoss} (A)}, we show the atom number per site $\langle \hat{n}_{\rm A} \rangle$ and the condensate fraction $|\psi|^2/\langle \hat{n}_{\rm A} \rangle$ as functions of the instantaneous lattice depth during the ramp-down dynamics.
We clearly see that in the strong dissipation region the onset of the atom loss or the order parameter growth shifts to the side of small lattice depth as $\gamma$ increases. 
This result suggests that one may experimentally observe the delay in the Mott-insulator melting by measuring the time evolution of the atom number and the momentum distribution during the ramp-down dynamics.

Having the above theoretical insights in mind, we perform the experiment for measuring ramp-down dynamics across the crossover from the Mott insulator to the superfluid.
The atom number and the momentum distribution during ramp-down dynamics are obtained from the fluorescence detection and the density distribution of the time-of-flight (TOF) absorption image, respectively.
Our experiment starts with ramping up the lattice to $V_0 = 20~E_R$ for preparation of the singly-occupied Mott insulator state. The atom number is  tuned to be small enough that no doubly-occupied site exists. Subsequently we ramp down the lattice with applying the PA laser. The lattice ramp-down speed is $-2 \, E_R$/ms. After ramping down the lattice to the final lattice depth, we perform the fluorescence detection for measuring the atom number, or we suddenly turn off all the trap and take the absorption image after 8 ms ballistic expansion for obtaining the density distribution.

We first focus on the atom loss measurement during ramp-down dynamics.
\textbf{Figure \ref{fig:AtomLoss} (B)} shows the atom number measured with various dissipation strengths.
The experimental result well reproduces the overall features of the calculation shown in \textbf{Fig. \ref{fig:AtomLoss} (A)} (left).
Specifically, the significant atom loss starts around $V_0 = 10~E_R$ in the presence of weak dissipation (\textcircled{\footnotesize 2}), while the atom number is conserved during ramping down the lattice without dissipation (\textcircled{\footnotesize 1}).
This onset shifts to the deep lattice side as $\gamma$ increases (\textcircled{\footnotesize 3}) for weak dissipation ($\gamma < 2$).
However, when $\gamma$ increases further from $\gamma \sim 2$, the onset shifts to the shallow lattice side (\textcircled{\footnotesize 4}). 
In order to identify the onset, we fit the double linear function to the data (\textbf{Fig. \ref{fig:AtomLoss} (C)}), which are shown in \textbf{Fig. \ref{fig:AtomLoss} (B)}.
In the presence of on-site dissipation, the atom loss is correlated with the melting of the Mott insulator in ramp-down dynamics because the melting creates the double occupation which is blasted out by the PA laser. Our result suggests that the melting of the Mott insulator is delayed for strong on-site dissipation. Quantitatively, the onset changes from $V_0=11.7(4)~E_R$ to $V_0=9.2(4)~E_R$ at the maximum as $\gamma$ increases. This corresponds to the increase of $zJ/U$ by a factor of 2.2. These behaviors capture the essence of the theoretical predictions mentioned in the above.

\textbf{Figure \ref{fig:AbsImage} (A)} shows a series of TOF absorption images obtained by changing the final lattice depth from Mott insulator regime to superfluid regime with various strength of the dissipation. 
Without dissipation, we clearly observe the transition from a Mott insulator state to a superfluid state as shown in \textbf{Fig. \ref{fig:AbsImage} (A)} of $\gamma$ = 0: in the deep lattice such as $V_0=20~E_R$, we obtain a broad distribution with no pattern, which indicates that the atoms have no phase coherence corresponding to the Mott insulator state. As ramping down the lattice across the critical depth of $V_0 =11.3~E_R$, which is calculated from the scattering length of ${}^1S_0$ state of $^{174}$Yb~\cite{PhysRevA.77.012719}, we obtain a clear interference pattern characterizing the presence of the phase coherence of the superfluid state. 
In the presence of dissipation, the observed transition is significantly modified, as shown in \textbf{Fig. \ref{fig:AbsImage} (A)}.
As the strength of the dissipation increases, the interference pattern becomes unclear in the shallow lattice regime. 
For strong dissipation such as $\gamma \sim 5$, any pattern cannot be observed.
This result indicates that the growth of the phase coherence is suppressed by the strong dissipation. 

In order to evaluate the phase coherence quantitatively, we introduce the visibility of the interference peaks as $v=(N_{\rm{max}}-N_{\rm{min}})/(N_{\rm{max}} +N_{\rm{min}})$~\cite{PhysRevA.72.053606}. Here, $N_{\rm{max}}$ is the sum of the number of atoms in the regions of first-order interference peaks and $N_{\rm{min}}$ is the sum of the number of atoms in the regions at the same distance from the central peak along the diagonals. 
While the visibility increases with the ramp-down of the lattice, this increase becomes more moderate in the stronger dissipation, as shown in \textbf{Fig. \ref{fig:AbsImage} (B)}. 
Especially, clear dependence on the strength of the dissipation is observed below the depth of  $V_0=11~E_R$, which is around the calculated critical depth at $\gamma = 0$. As shown in \textbf{Fig. \ref{fig:AbsImage} (B)}, the effect of the dissipation on the width of the crossover region is observed as more moderate growing of the visibility below the depth of $11~E_R$ associated with the increase of $\gamma$. In addition, as shown in \textbf{Fig. \ref{fig:AbsImage} (C)}, the dissipation moderates the narrowing of the width (in $\mu$m) of the density distribution and the slope of the width (in $\mu$m) with respect to the lattice depth becomes less steep as $\gamma$ increases. 
Narrowing the width of the distribution indicates the localization of the state in the momentum space. 
All of these measurements support the delay in the melting of the singly-occupied Mott insulator in the ramp-down dynamics as an effect of the on-site dissipation as we see in the calculation of the condensate fraction shown in \textbf{Fig. \ref{fig:AtomLoss}}. Notice that the observation of the excitation gap, which is the direct evidence of the formation of the Mott insulator, is difficult in this dissipative system because the excitation spectrum should have a broad linewidth determined by the inelastic collision rate $\Gamma_{\rm PA}$ of a few tens of kHz.

\subsection*{Quenching the dissipation}

It is important to experimentally check whether this behavior is attributed to some heating effect by the PA laser.
For this purpose, we measure the phase coherence after turning off the PA laser.
If the absence of the interference pattern is attributed to the heating, the phase coherence is no longer restored after the PA laser is turned off.
In contrast, if the state after the ramp-down of the lattice is still a Mott insulator, the phase coherence can be restored. 
Similarly to the measurement of \textbf{Fig. \ref{fig:AbsImage} (A)}, we ramp down the lattice to a final lattice depth  in $-2\,E_R$/ms with the maximum strength of dissipation $\gamma$ = 4.6(4).
Then, we suddenly turn off the PA laser and investigate the subsequent time evolution of the atoms in the lattice by observing the phase coherence through a TOF absorption image at some hold time.

\begin{figure}[tbp]
\includegraphics[width=11cm,clip]{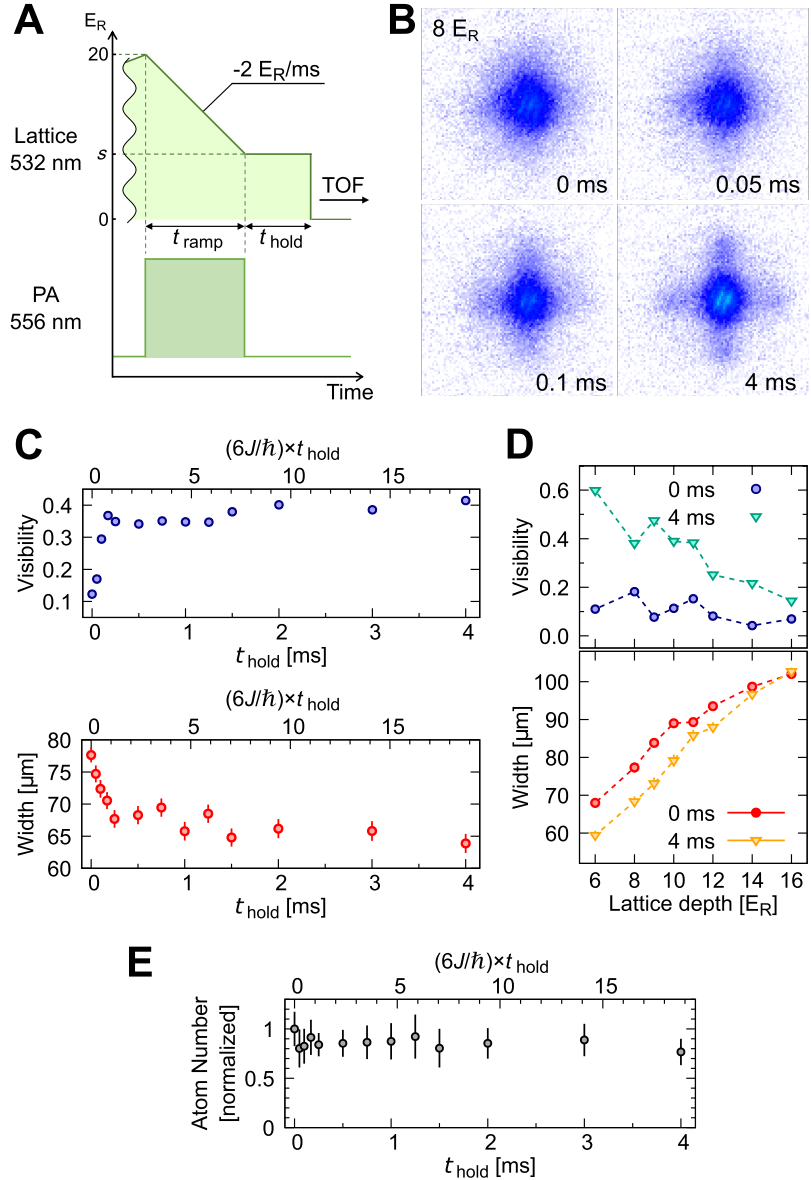}
\caption{
{\bf Dynamics after turning off the dissipation.} 
\textbf{(A)}, Experimental sequence for the observation of the dynamics after turning off the dissipation. After ramping up the lattice to $V_0 =20~E_R$ for preparing the Mott insulator state, we ramp down the lattice to the final lattice depth $V_0 =s ~E_R$ with applying the PA laser. 
The ramp-down speed is $-2 E_R/\rm{ms}$ and the ramp-down time is $t_{\rm{ramp}}=(20-s)/2$ ms. After ramping down the lattice, we turn off the PA laser and hold the lattice for $t_{\rm{hold}}$. 
\textbf{(B)}, Time evolution of TOF image after turning off the dissipation. 
The hold time after turning off the PA laser is shown at the bottom right of each image. 
\textbf{(C)}, Time evolution of the visibility and the width. 
\textbf{(D)}, Lattice depth dependence of the visibility and the width with 0 ms and 4 ms hold time. 
\textbf{(E)}, Atom number during after turning off the dissipation. The atom number is normalized by the initial atom number at $t_{\rm hold}$ = 0 ms. 
}
\label{fig:Quench}
\end{figure}

The result for the case of the final lattice of $V_0 =8~E_R$ is shown in \textbf{Fig. \ref{fig:Quench} (B)} for the observed TOF images and \textbf{Fig. \ref{fig:Quench} (C)} for the evolution of the visibility and width of the density distribution. 
After some hold time, an interference pattern grows. It serves as a direct signature of the restoration of the phase coherence, indicating that the absence of the interference pattern in \textbf{Fig. \ref{fig:AbsImage} (A)} is not completely attributed to the heating.
We confirm that the total atom number is conserved in this dynamics as shown in \textbf{Fig. \ref{fig:Quench} (E)}. This means that the evaporative cooling which could possibly explain the observed behavior does not occur during the dynamics. Importantly, the atoms after turning off the dissipation can be considered as an isolated (closed) system. Therefore, the observed dynamics in our experiment should not be considered as the usual thermal relaxation with the environment, but the relaxation in the isolated quantum system, which is a hot topic actively studied in recent experiments and theories~\cite{Polkovnikov-11,Langen-16,Eisert-15,Nandkishore-15}. Here, we consider the tunneling time as a relevant time scale because the superfluid state is realized through the process of delocalization of the particles by the tunneling. As shown in \textbf{Fig. \ref{fig:Quench} (C)}, the time constant of the increase of the visibility and the decrease of the width is comparable to the tunneling time $(6J/\hbar)^{-1}$ = 0.21 ms. \textbf{Figure \ref{fig:Quench} (D)} shows the visibility and the width of the density distribution, 0 ms and 4 ms after the ramp-down for various final lattice depths, similarly indicating the restoration of the coherence.

\section{Discussion}

We have realized the engineered dissipative Bose-Hubbard system by introducing a controllable strength of two-body inelastic collision with use of a PA laser. By exploiting the highly controllable nature of the dissipation, we have investigated the effect of the dissipation on the quantum phase transition from the Mott insulator state to the superfluid state in the lattice ramp-down dynamics. We have observed that the melting of the Mott state is delayed and the growth of the phase coherence is suppressed for the strong on-site dissipation. The favored state depends on the type of the dissipation. For example, the stabilization of the superfulid state with use of a well-designed off-site dissipation is proposed~\cite{Diehl-08}. Thanks to the dramatic change in the onset of the Mott-insulator melting, as shown in the fact of increase of $zJ/U$, we can access the interesting problem of quenching the dissipation across the crossover from the Mott insulator to the superfluid, where turning off the dissipation corresponds to a sudden parameter change of the Bose-Hubbard system~\cite{Altman-02}. In this method, the required time for turning off the dissipation could be very short while the sudden change of the depth of optical lattice needs a certain time in order to stabilize the power of the lattice laser as well as to prevent non-adiabatic inter-band transition. Moreover, while we have used $^{174}$Yb, which is a bosonic isotope of an alkali-earth-like species, to demonstrate our method for controlling the dissipation, it is applicable generally to other atomic species that can be coupled to a state of lossy PA molecule. The crossover properties can also be caused by varying the on-site interaction (see Fig. S10 (B) of the Supplementary Materials). Controlling the on-site interaction with Feshbach resonance, for example using alkali atoms, enables us to investigate wider range of strength of dissipation including infinitely strong, since weakening the on-site interaction corresponds to strengthen the dissipation $\gamma$. Our work opens a new way to study the quantum many-body system by controlling the dissipation.

\section{Materials and Methods}

\subsection*{Preparation of ultracold ${}^{174}$Yb atoms in an optical lattice}

After collecting atoms with a magneto-optical trap (MOT) with the intercombination transition of ${}^1S_0 \leftrightarrow {}^3P_1$, we load the atoms into a crossed far-off-resonant trap (FORT). 
Subsequently, an evaporative cooling is performed, resulting in an almost pure BEC with no discernible thermal component. 
The atom number of BEC is controlled by changing the collecting time of atoms in MOT. 
The trap frequencies of the FORT at the final stage of the evaporative cooling are $(\omega_{x'}, \omega_{y'}, \omega_z)/2\pi =(162, 31,166)$ Hz. Here, the $x'$- and $y'$-axes are tilted from the lattice axes ($x$ and $y$) by 45$^{\circ}$ in the same plane.
Then, the BEC is loaded into a 3D optical lattice  with the wavelength of the lattice laser $\lambda_L$ of 532 nm. 
A typical atom number of BEC is $6 \times 10^4$ in a measurement of the photo-association rate of doubly-occupied sites, or $1 \times 10^4$ in a measurement of two-body loss rate and observation of the quantum phase transition for atoms without multiply-occupied sites. 

\subsection*{Detail of PA}
By applying the PA laser to the atoms in the lattice, two atoms in the doubly-occupied sites are photoassociated into the ${}^1S_0 + {}^3P_1$ molecular state whose vibrational quantum number $v'_e=16$.
These molecules are immediately dissociated into the two ground state atoms. 
We note that the photon scattering of the atoms due to the PA laser can be negligible because the selected PA line is 3.7 GHz below the atomic transition line, which is about $2\times 10^4$ times larger than its natural line width.
$\beta_{\rm{PA}}$ for various intensities is determined through the loss dynamics of the atoms by measuring the remaining atom number  with the fluorescence detection method after applying the PA laser to the atoms in the lattice with depth of $V_0 =14~E_R$.
In this lattice depth, the system is in a state with singly- and doubly-occupied sites.
After the PA laser is applied, the remaining atom number $N(t)$ decreases as $N(t)=N_1 + N_2 \exp(-\Gamma_{\rm{PA}} t)$, where $N_1$ and $N_2$ are the initial atom number in the singly and doubly-occupied site respectively.

\section*{Acknowledgements}
We thank Y. Ashida, M. Barbier and P. Naidon for fruitful discussions. 
This work was supported by MEXT/JSPS KAKENHI Grant Numbers JP25220711, JP26247064, JP16H00990, JP16H01053, JP16H00801, CREST, JST No. JPMJCR1673, and Matsuo Foundation. 
T.T. acknowledges support from the Japan Society for the Promotion of Science (KAKENHI Grant Number JP16J01590). 

\section*{Author contributions}
T. Tomita, S. Nakajima and Y. Takasu carried out experiments and the data analysis. 
I. Danshita carried out the theoretical calculation. Y. Takahashi conducted the whole experiment. All the authors contributed to the writing of the manuscript. 

\section*{Competing interests}
The authors declare no competing financial interests. 

\section*{Data and materials availability}
All data needed to evaluate the conclusions in the paper are present in the paper and/or the Supplementary Materials.
Additional data related to this paper may be requested from T.T. (tomita@scphys.kyoto-u.ac.jp).

%


\clearpage
\begin{center}
\noindent\textbf{\large{Supplementary Materials for ``Observation of the Mott Insulator to Superfluid Crossover of a Driven-Dissipative Bose-Hubbard System"}}
\\\bigskip
Takafumi Tomita$^{1}$, Shuta Nakajima$^{1}$, Ippei Danshita$^{2}$, Yosuke Takasu$^{1}$, and Yoshiro Takahashi$^{1}$
\\\vspace{0.1cm}
\small{$^{1}$\emph{Department of Physics, Kyoto University, Kyoto 606-8502, Japan}}\\
\small{$^{2}$\emph{Yukawa Institute for Theoretical Physics, Kyoto University, Kyoto 606-8502, Japan}}\\
\end{center}
\bigskip

\setcounter{section}{0}
\setcounter{equation}{0}
\setcounter{figure}{0}

\renewcommand{\thesection}{S\arabic{section}}
\renewcommand{\figurename}{\textbf{Fig.}}
\renewcommand{\thefigure}{\textbf{S\arabic{figure}}}
\makeatletter
	\renewcommand{\theequation}{\thesection.\arabic{equation}}
	\@addtoreset{equation}{section}
\makeatother

\section{Derivation of the dissipative Bose-Hubbard model}
In this section, we present a detailed derivation of the dissipative Bose-Hubbard model with a two-body loss term from the coupled atom-molecule mixture model with a one-body molecular loss term. The derivation is based on a kind of perturbative approach developed in Ref.~\cite{garcia-09}.
\subsection{Models}

We start with the Markovian master equation for the coupled atom-molecule mixture model with a one-body molecular loss term,
\begin{eqnarray}
\hbar\frac{d}{dt}\hat{\rho} = - i \left[\hat{H}, \hat{\rho}\right] 
+ L_{\rm M}(\hat{\rho}).
\label{eq:master_init}
\end{eqnarray}
The Hamiltonian is given by
\begin{eqnarray}
\hat{H} = \hat{H}_0 + \hat{H}_{\rm am} + \hat{H}_{\rm hop},
\label{eq:hamil_AM}
\end{eqnarray}
where
\begin{eqnarray}
\hat{H}_0 &=& \sum_j \left( D \hat{n}_{{\rm M},j} + \frac{U}{2} \hat{n}_{{\rm A},j}(\hat{n}_{{\rm A},j}-1) 
+ W \hat{n}_{{\rm A},j}\hat{n}_{{\rm M},j}\right),
\\
\hat{H}_{\rm am} &=& \sum_{j}g\left( \hat{m}_{j}^{\dagger} \hat{a}_j \hat{a}_j + {\rm h.c.} \right),
\\
\hat{H}_{\rm hop} &=& 
-\sum_{\langle j,k \rangle} J\left( \hat{a}_j^{\dagger} \hat{a}_k + {\rm h.c.} \right).
\end{eqnarray}
The quantum phases of this atom-molecular Hamiltonian have been theoretically studied in Ref.~\cite{rousseau-08}.
We anticipate that the molecule consists of a ${}^1S_0$ atom and a ${}^3 P_1$ atom as in the experiment such that its linewidth is on the order of 1 MHz due to the short lifetime of the latter state of atom. Hence, we have to include the one-body loss term of molecules in the master equation,
\begin{eqnarray}
L_{\rm M}(\hat{\rho}) = \frac{\hbar\Gamma_{\rm M}}{2} \sum_j \left(
- \hat{n}_{{\rm M},j} \hat{\rho} - \hat{\rho} \hat{n}_{{\rm M},j} + 2\hat{m}_j \hat{\rho} \hat{m}_j^{\dagger}\right). 
\end{eqnarray}
Equation (\ref{eq:master_init}) describes the dynamics of ultracold bosonic atoms in an optical lattice coupled with a molecular state via photo-association (PA) laser.
$\hat{a}_{j}$ and $\hat{m}_j$ annihilate an atom and a molecule on site $j$ while $\hat{n}_{{\rm A},j} = \hat{a}^{\dagger}_j \hat{a}_j$ and $\hat{n}_{{\rm M},j} = \hat{m}^{\dagger}_j \hat{m}_j$ are the density operators of atoms and molecules.
$\Gamma_{\rm M}$, $D$, $U$, $W$, $g$, and $J$ denote the one-body loss of molecules, the detuning of the PA coupling from the molecular state, the on-site interaction between two atoms, the on-site interaction between an atom and a molecule, the atom-molecule coupling, and the hopping of atoms. $\langle j,k\rangle$ represents nearest-neighboring pairs of lattice sites. Since $\hbar\Gamma_{\rm M}\gg \max(|D|,|g|,|J|)$, a molecule created on a lattice site via the PA laser decays much earlier than the creation of another molecule on the same site. In this sense, we can safely assume the hardcore constraint on the molecules, which forbids more than one molecules to occupy a single site.
We aim to show that when $\hbar\Gamma_{\rm M}\gg \max\left( |D|, |U|, |W|, |g|, |J| \right)$, the molecular degrees of freedom can be properly projected out by means of a perturbation theory~\cite{garcia-09}.
such that the system is well approximated by the following effective master equation,
\begin{eqnarray}
\hbar\frac{d}{dt}\hat{\rho}_{\rm eff} = -i \left[ 
\hat{H}_{\rm eff}, \hat{\rho}_{\rm eff}
\right]
+ L_2(\hat{\rho}_{\rm eff}),
\label{eq:master_BH}
\end{eqnarray}
where
\begin{eqnarray}
\hat{H}_{\rm eff} = \sum_j \frac{U}{2} \hat{n}_{{\rm A},j}(\hat{n}_{{\rm A},j}-1) 
-\sum_{\langle j,k \rangle} J\left( \hat{a}_j^{\dagger} \hat{a}_k + {\rm h.c.} \right),
\end{eqnarray}
\begin{eqnarray}
L_2(\hat{\rho}_{\rm eff}) = \frac{\hbar\Gamma_{\rm PA}}{4}\sum_j \left(
- \hat{a}_j^{\dagger}\hat{a}_j^{\dagger}\hat{a}_j\hat{a}_j \hat{\rho}_{\rm eff}
- \hat{\rho}_{\rm eff} \hat{a}_j^{\dagger}\hat{a}_j^{\dagger}\hat{a}_j\hat{a}_j 
+ 2 \hat{a}_j\hat{a}_j \hat{\rho}_{\rm eff}\hat{a}_j^{\dagger}\hat{a}_j^{\dagger}
\right),
\end{eqnarray}
and 
\begin{eqnarray}
\Gamma_{\rm PA} = 8\frac{g^2}{\hbar^2 \Gamma_{\rm M}}.
\label{eq:gamma_PA}
\end{eqnarray}
Equations (\ref{eq:master_BH})-(\ref{eq:gamma_PA}) are written also in the main text as Eqs.~(1)-(4).

In order to express the density matrix of the system more explicitly, we define the local Fock state on site $j$,
\begin{eqnarray}
|n_a,n_m\rangle_j = \frac{1}{\sqrt{n_a ! \, n_m !}}
 (\hat{a}^{\dagger}_j)^{n_a}(\hat{m}^{\dagger}_j)^{n_m}|vac\rangle_j
\end{eqnarray}
We set the maximum number of atoms per site to be $d-1$. Hence, the dimension of the local Hilbert space is $2d$. While the maximum number of bosonic atoms per site is in principle the total number of atoms, at a finite filling factor the occupation probability of large-$n_a$ states decays exponentially. This means that setting the cutoff of the local Hilbert space at $n_a = d-1$ does not affect results of numerical calculations in practice as long as $U>0$ and $d$ is sufficiently large~\cite{kuhner-98}.  

For convenience, we introduce a simpler notation for the local state,
\begin{eqnarray}
|l\rangle_j = \left\{
\begin{array}{cc}
|n_a = l -1, n_m=0\rangle_j, & {\rm when}\,\, 1 \leq l \leq d, \\
|n_a = l - 1 - d, n_m = 1 \rangle_j, &  {\rm when}\,\, d+1 \leq l \leq 2d.
\end{array}
\right.
\end{eqnarray}
Using the local states defined above, we express a general form of the density matrix as
\begin{eqnarray}
\hat{\rho} = \sum_{l_1, l_2, \ldots}\sum_{m_1,m_2, \ldots} 
\rho_{l_1,l_2, \ldots}^{m_1,m_2, \ldots} \prod_{j} |l_j \rangle \langle m_j|_j.
\end{eqnarray}

Regarding the density matrix $\hat{\rho}$, which is a $(2d)^M\times (2d)^M$ matrix, as a $(2d)^{2M}$-dimensional vector ${\boldsymbol \rho}$, we can rewrite the master equation in the following form,
\begin{eqnarray}
\hbar\frac{d}{dt} {\boldsymbol \rho} = \left( \hat{\mathcal{M}}_0
+ \hat{\mathcal{V}}_{\rm am} + \hat{\mathcal{V}}_{\rm hop}
\right){\boldsymbol \rho},
\label{eq:master_SO}
\end{eqnarray}
where the superoperators $\hat{\mathcal{M}}_0$, $\hat{\mathcal{V}}_{\rm am}$, and $\hat{\mathcal{V}}_{\rm hop}$ are $(2d)^{2M}\times (2d)^{2M}$ matrices originated from $\hat{H}_0$ and $L_{\rm M}(\hat{\rho})$, $\hat{H}_{\rm am}$, and $\hat{H}_{\rm hop}$, respectively. 
Associated with the change of the notation from $\hat{\rho}$ to ${\boldsymbol \rho}$, we also express the local matrix as the following vector,
\begin{eqnarray}
|l_j,m_j)_j = |l_j\rangle \langle m_j|_j.
\end{eqnarray}
This vector satisfies the orthonormality condition,
\begin{eqnarray}
(l_j',m_j'|l_j,m_j) = \delta_{l'_j,l_j}\delta_{m_j',m_j}.
\end{eqnarray}
%
\subsection{Local projection}
In order to derive the effective master equation (\ref{eq:master_BH}), we need to express some of the superoperators explicitly and introduce the projection superoperator. The non-perturbative superoperator $\hat{\mathcal{M}}_0$ can be expressed as a sum of commuting local superoperators $\hat{\mathcal{M}}^{\rm loc}_{0,j}$,
\begin{eqnarray}
\hat{\mathcal{M}}_0 = \sum_j \hat{\mathcal{M}}^{\rm loc}_{0,j},
\end{eqnarray}
where
\begin{eqnarray}
\hat{\mathcal{M}}^{\rm loc}_{0} = \sum_{l=1}^{2d}\sum_{m=1}^{2d} |l,m) (l,m|
\,  i \left(\left(E_l^{(0)}\right)^{\ast} - E_m^{(0)}\right)
+ \sum_{l=1}^{d}\sum_{m=1}^{d}|l,m)(d+l,d+m|\hbar\Gamma_{\rm M},
\label{eq:M0_loc}
\end{eqnarray}
and
\begin{eqnarray}
E_l^{(0)} = \left\{
\begin{array}{cc}
 \frac{U}{2}(l-1)(l-2), & {\rm when}\,\, 1\leq l \leq d,
 \\
 D + \frac{U}{2}(l-d-1)(l-d-2) + W(l-d-1) - i\frac{\hbar\Gamma_{\rm M}}{2}, &
 {\rm when}\,\, d+1\leq l \leq 2d.
\end{array}
\right.
\end{eqnarray}
The perturbative superoperator $\hat{\mathcal{V}}_{\rm am}$ originated from the atom-molecule coupling $\hat{H}_{\rm am}$ can be also expressed as a sum of commuting local superoperators,
\begin{eqnarray}
\hat{\mathcal{V}}_{\rm am} = \sum_j \hat{\mathcal{V}}^{\rm loc}_{{\rm am},j}
\end{eqnarray}
where
\begin{eqnarray}
\hat{\mathcal{V}}_{\rm am}^{\rm loc} &=& 
\sum_{l=1}^{2d}\sum_{m=1}^{d}\left(|l,m)(l,d+m-2| + |l,d+m-2)(l,m| \right)\tilde{g}_m 
\nonumber \\
&& - \sum_{l=1}^{d}\sum_{m=1}^{2d}\left( |l,m)(d+l-2,m| + |d+l-2,m)(l,m| \right)\tilde{g}_l
\label{eq:Vam_loc}
\end{eqnarray}
and
\begin{eqnarray}
\tilde{g}_l = g\sqrt{(l-1)(l-2)}.
\end{eqnarray}
We omitted the site index $j$ of $\hat{\mathcal{M}}^{\rm loc}_{0,j}$ and $\hat{\mathcal{V}}^{\rm loc}_{{\rm am},j}$ in Eqs.~(\ref{eq:M0_loc}) and (\ref{eq:Vam_loc}) because they do not depend on $j$.

We do not write an explicit expression of the superoperator $\hat{\mathcal{V}}_{\rm hop}$ because it is unnecessary for our purpose. Nevertheless, it is worth noting that $\hat{\mathcal{V}}_{\rm hop}$ changes neither the number of atoms nor the number of molecules. This means that this superoperator does not have matrix elements connecting the effective Hilbert space with the truncated one.

In order to construct the projection superoperator, we need to solve the following eigenvalue problem,
\begin{eqnarray}
\hat{\mathcal{M}}_0^{\rm loc} |v_\alpha) &=& \lambda_{\alpha} |v_\alpha),
 \\
(w_\alpha|\hat{\mathcal{M}}_0^{\rm loc}  &=& (w_\alpha| \lambda_{\alpha}.
\end{eqnarray}
Notice that the left eigenvector $(w_\alpha|$ in general is not equal to the conjugate of the right eigenvector $|v_\alpha)$ because $\hat{\mathcal{M}}_0^{\rm loc}$ is not Hermitian. The eigenvectors satisfy the following orthonormality condition,
\begin{eqnarray}
(w_{\alpha'} | v_{\alpha}) = \delta_{\alpha,\alpha'}.
\end{eqnarray}
When we derive the effective model, we utilize the fact that $d^2$ eigenvalues have the property $|\lambda_{\alpha}|=O(\max(|D|,|U|,|W|))$ and the other eigenvalues have $|\lambda_{\alpha}|=O(\hbar\Gamma_{\rm M})$. The local subspaces that include states with the former and latter properties are denoted by $\mathcal{D}_{\rm eff}^{\rm loc}$ and $\mathcal{D}_{\rm trc}^{\rm loc}$, respectively. 

We define the local projection superoperators as
\begin{eqnarray}
\hat{\mathcal{P}}_{\alpha}^{\rm loc} = |v_{\alpha})(w_{\alpha}|.
\end{eqnarray}
This superoperator projects a state or a superoperator on state $\alpha$.
From these projectors, we construct the projection superoperator on the effective Hilbert space,
\begin{eqnarray}
\hat{\mathcal{P}}_{\rm eff}^{\rm loc} &=& 
\sum_{\alpha \in \mathcal{D}_{\rm eff}^{\rm loc}}  \hat{\mathcal{P}}_{\alpha}^{\rm loc} 
\nonumber \\
&=& 
\sum_{l=1}^d \sum_{m=1}^d \left( |l,m) (l,m| + |l,m)(d+l,d+m| \right).
\label{eq:Peff_loc}
\end{eqnarray}
Notice that from the first line to the second line of Eq.~(\ref{eq:Peff_loc}), we neglected the terms on the order of $\max(|D|,|U|,|W|)/(\hbar\Gamma_{\rm M})$ on the basis of the assumption that $\hbar\Gamma_{\rm M} \gg \max(|D|,|U|,|W|)$.

\subsection{Second-order perturbation}
From the local projection operator of Eq.~(\ref{eq:Peff_loc}), we construct the projection operators for the entire system as
\begin{eqnarray}
\hat{\mathcal{P}}_{\rm eff} = \prod_{j=1}^M \hat{\mathcal{P}}_{{\rm eff},j}^{\rm loc},
\\
\hat{\mathcal{P}}_{\rm trc} = \hat{\mathcal{I}} - \hat{\mathcal{P}}_{\rm eff},
\end{eqnarray}
where $\hat{\mathcal{I}}$ is the global identity matrix.
Multiplying $\hat{\mathcal{P}}_{\rm eff}$ on Eq.~(\ref{eq:master_SO}) from the left and using the facts that $\hat{\mathcal{P}}_{\rm eff} \hat{\mathcal{V}}_{\rm am}\hat{\mathcal{P}}_{\rm eff} = 0$ and $\hat{\mathcal{P}}_{\rm eff} \hat{\mathcal{V}}_{\rm hop}\hat{\mathcal{P}}_{\rm trc} = 0$, we obtain
\begin{eqnarray}
\hbar\frac{d}{dt}{\boldsymbol \rho}_{\rm eff} = 
\hat{\mathcal{P}}_{\rm eff}
\left(
\hat{\mathcal{M}}_0 + \hat{\mathcal{V}}_{\rm hop}
\right)
\hat{\mathcal{P}}_{\rm eff}{\boldsymbol \rho}_{\rm eff} 
+ \hat{\mathcal{P}}_{\rm eff}\hat{\mathcal{V}}_{\rm am}\hat{\mathcal{P}}_{\rm trc}
{\boldsymbol \rho}_{\rm trc},
\label{eq:rho_eff_temp}
\end{eqnarray}
where
\begin{eqnarray}
{\boldsymbol \rho}_{\rm eff} &=& \hat{\mathcal{P}}_{\rm eff} {\boldsymbol \rho},
\\
{\boldsymbol \rho}_{\rm trc} &=& \hat{\mathcal{P}}_{\rm trc} {\boldsymbol \rho}.
\end{eqnarray}
In Eq.~(\ref{eq:rho_eff_temp}), it is obvious that the first and second terms in the right hand side already agree with the term $-i\left[\hat{H}_{\rm eff}, \hat{\rho}_{\rm eff}\right]$ in Eq.~(\ref{eq:master_BH}).

We will next derive $L_2(\hat{\rho}_{\rm eff})$ in Eq.~(\ref{eq:master_BH}) from the last term in the right hand side of Eq.~(\ref{eq:rho_eff_temp}).
The components in $\hat{\mathcal{P}}_{\rm trc}$ that can give finite contributions to $\hat{\mathcal{P}}_{\rm eff}\hat{\mathcal{V}}_{\rm am}\hat{\mathcal{P}}_{\rm trc}$ are the ones written as
\begin{eqnarray}
\hat{\mathcal{P}}_{\beta} = \sum_{k} \hat{\mathcal{P}}_{\beta}^{(k)},
\end{eqnarray}
where
\begin{eqnarray}
\hat{\mathcal{P}}_{\beta}^{(k)} = \hat{\mathcal{P}}_{\beta,k}^{\rm loc}
\prod_{j\neq k} \hat{\mathcal{P}}_{{\rm eff},j}^{\rm loc},
\end{eqnarray}
and $\beta \in \mathcal{D}_{\rm trc}^{\rm loc}$.
Hence, Eq.~(\ref{eq:rho_eff_temp}) can be rewritten as
\begin{eqnarray}
\hbar\frac{d}{dt}{\boldsymbol \rho}_{\rm eff} = 
\hat{\mathcal{P}}_{\rm eff}
\left(
\hat{\mathcal{M}}_0 + \hat{\mathcal{V}}_{\rm hop}
\right)
\hat{\mathcal{P}}_{\rm eff}{\boldsymbol \rho}_{\rm eff} 
+ \hat{\mathcal{P}}_{\rm eff}\hat{\mathcal{V}}_{\rm am}
\sum_{\beta}\hat{\mathcal{P}}_{\beta}{\boldsymbol \rho}_{\beta}.
\label{eq:rho_eff}
\end{eqnarray}

In order for Eq.~(\ref{eq:rho_eff}) to be closed within the effective Hilbert space, we need to express ${\boldsymbol \rho}_{\beta}$ in terms of ${\boldsymbol \rho}_{\rm eff}$. 
For this purpose, we look into the equation for ${\boldsymbol \rho}_{\beta}$ given by
\begin{eqnarray}
\hbar \frac{d}{dt}{\boldsymbol \rho}_{\beta} = 
\left( \hat{\mathcal{R}} + \lambda_{\beta} \right) {\boldsymbol \rho}_{\beta}
+ \hat{P}_{\beta} \hat{\mathcal{V}}_{\rm am}\hat{P}_{\rm eff}{\boldsymbol \rho}_{\rm eff} + \hat{P}_{\beta}\hat{\mathcal{V}}\hat{P}_{\rm trc}{\boldsymbol \rho}_{\rm trc},
\label{eq:rho_beta}
\end{eqnarray}
where
\begin{eqnarray}
\hat{\mathcal{R}} &=& \sum_{k} \hat{\mathcal{P}'}_{\rm eff}^{(k)} \hat{\mathcal{M}'}_0^{(k)},
\\
\hat{\mathcal{M}'}_0^{(k)} &=& \sum_{j\neq k}  \hat{\mathcal{M}}_{0,j}^{\rm loc},
\\
\hat{\mathcal{P}'}_{\rm eff}^{(k)} &=& \hat{\mathcal{I}}_{k}^{\rm loc} \prod_{j\neq k}
\hat{\mathcal{P}}_{{\rm eff},j}^{\rm loc}.
\end{eqnarray}
We neglect the last term in the right hand side of Eq.~(\ref{eq:rho_beta}) because it gives higher-order contributions with respect to $\hat{\mathcal{V}}$.
Making a variable transformation,
\begin{eqnarray}
{\boldsymbol \rho}_{\beta}(t) &=& e^{\hat{\mathcal{R}}t/ \hbar}
\tilde{\boldsymbol \rho}_{\beta}(t),
\\
{\boldsymbol \rho}_{\rm eff}(t) &=& e^{\hat{\mathcal{R}}t/ \hbar}\tilde{\boldsymbol \rho}_{\rm eff}(t),
\end{eqnarray}
Eq.~(\ref{eq:rho_beta}) is simplified a little,
\begin{eqnarray}
\hbar \frac{d}{dt}\tilde{\boldsymbol \rho}_{\beta} = 
\lambda_{\beta} \tilde{\boldsymbol \rho}_{\beta}
+ \hat{P}_{\beta} \hat{\mathcal{V}}_{\rm am}\hat{P}_{\rm eff}\tilde{\boldsymbol \rho}_{\rm eff}.
\label{eq:rho_beta_tilde}
\end{eqnarray}
Formally solving Eq.~(\ref{eq:rho_beta_tilde}), we obtain
\begin{eqnarray}
\tilde{\boldsymbol \rho}_{\beta}(t) = \frac{1}{\hbar} e^{\lambda_{\beta} t/ \hbar} 
\int_0^t d\tau 
e^{-\lambda_{\beta}\tau/ \hbar}\hat{P}_{\beta} \hat{\mathcal{V}}_{\rm am}\hat{P}_{\rm eff}\tilde{\boldsymbol \rho}_{\rm eff}(\tau).
\end{eqnarray}
Performing a partial integral, this solution becomes
\begin{eqnarray}
\tilde{\boldsymbol \rho}_{\beta}(t) = 
- \frac{1}{\lambda_{\beta}}
\hat{\mathcal{P}}_{\beta} \hat{\mathcal{V}}\hat{\mathcal{P}}_{\rm eff}
\left(
\tilde{\boldsymbol \rho}_{\rm eff}(t) 
- e^{\lambda_{\beta}t  / \hbar}\tilde{\boldsymbol \rho}_{\rm eff}(0)
\right)
+ \frac{e^{\lambda_{\beta}t/ \hbar}}{\lambda_{\beta}}
\int_{0}^{t}d\tau e^{-\lambda_{\beta} \tau / \hbar} \hat{\mathcal{P}}_{\beta} \hat{\mathcal{V}}\hat{\mathcal{P}}_{\rm eff} \frac{d}{d\tau} \tilde{\boldsymbol \rho}_{\rm eff}(\tau).
\label{eq:formal_sol}
\end{eqnarray}
The remaining integral can be neglected because it is of higher order in $\max(|D|,|U|,|W|,|g|,|J|)/(\hbar\Gamma_{\rm M})$. The second term in Eq.~(\ref{eq:formal_sol}), which includes $e^{\lambda_{\beta}t/ \hbar}\tilde{\boldsymbol \rho}_{\rm eff}(0)$, decays very quickly on the order of $1/\Gamma_{\rm M}$ so that it can be also neglected as long as we are interested in much longer time scale than $1/\Gamma_{\rm M}$. Moreover, $\lambda_{\beta} = - \frac{\hbar\Gamma_{\rm M}}{2}$ in its leading order. Hence, ${\boldsymbol \rho}_{\beta}$ is well approximated as
\begin{eqnarray}
{\boldsymbol \rho}_{\beta}(t) = 
\frac{2}{\hbar \Gamma_{\rm M}}
\hat{\mathcal{P}}_{\beta} \hat{\mathcal{V}}\hat{\mathcal{P}}_{\rm eff}
{\boldsymbol \rho}_{\rm eff}(t).
\label{eq:rho_beta_final}
\end{eqnarray}

Substituting Eq.~(\ref{eq:rho_beta_final}) into Eq.~(\ref{eq:rho_eff}), we obtain 
\begin{eqnarray}
\hbar \frac{d}{dt}{\boldsymbol \rho}_{\rm eff} = 
\hat{\mathcal{P}}_{\rm eff}
(\hat{\mathcal{M}}_0 + \hat{\mathcal{V}}_{\rm hop})\hat{\mathcal{P}}_{\rm eff}{\boldsymbol \rho}_{\rm eff} 
+ \frac{2}{\hbar \Gamma_{\rm M}}
\sum_{\beta\in \mathcal{D}_{\rm trc}^{\rm loc}}
\hat{\mathcal{P}}_{\rm eff}\hat{\mathcal{V}}_{\rm am}\hat{\mathcal{P}}_{\beta}
\hat{\mathcal{V}}_{\rm am}\hat{\mathcal{P}}_{\rm eff}{\boldsymbol \rho}_{\rm eff},
\label{eq:rho_eff_final}
\end{eqnarray}
where
\begin{eqnarray}
\sum_{\beta} \hat{\mathcal{P}}_{\rm eff}\hat{\mathcal{V}}_{\rm am} \hat{\mathcal{P}}_{\beta} \hat{\mathcal{V}}_{\rm am} \hat{\mathcal{P}}_{\rm eff}= 
\sum_{k} \sum_{\beta} \hat{\mathcal{P}}_{{\rm eff},k}^{\rm loc}\hat{\mathcal{V}}_{{\rm am},k}^{\rm loc} \hat{\mathcal{P}}_{\beta,k}^{\rm loc} \hat{\mathcal{V}}_{{\rm am},k}^{\rm loc} \hat{\mathcal{P}}_{{\rm eff},k}^{\rm loc}
\end{eqnarray}
and
\begin{eqnarray}
\hat{\mathcal{P}}_{{\rm eff}}^{\rm loc}\hat{\mathcal{V}}_{{\rm am}}^{\rm loc} \hat{\mathcal{P}}_{\beta}^{\rm loc} \hat{\mathcal{V}}_{{\rm am}}^{\rm loc} \hat{\mathcal{P}}_{{\rm eff}}^{\rm loc} &=& 
\sum_{l=1}^{d}\sum_{m=1}^{d}
\biggl(
|l,m)(l,m|(\tilde{g}_m^2 + \tilde{g}_l^2)
- |l,m)(l+2,m+2|2\tilde{g}_{l+2}\tilde{g}_{m+2}
\nonumber \\
&& + |l,m)(d+l,d+m|(\tilde{g}_m^2 + \tilde{g}_l^2)
- |l,m)(d+l+2,d+m+2|2\tilde{g}_{l+2}\tilde{g}_{m+2}
\biggr).
\end{eqnarray}
Rewriting Eq.~(\ref{eq:master_BH}) with use of superoperators and the vector form of the density matrix, and setting $\Gamma_{\rm PA}=8g^2/(\hbar^2 \Gamma_{\rm M})$, we recognize that the derived effective master equation, which is Eq.~(\ref{eq:rho_eff_final}), is equivalent to that for the dissipative Bose-Hubbard model with the two-body loss term. Thus, we have successfully derived the dissipative Bose-Hubbard model from the coupled atom-molecule mixture model with the one-body molecular loss term. 

\section{Loss dynamics from the Mott insulating state with double filling}
In this section, we analyze the dynamics of the Mott insulator with two bosonic atoms per site in an optical lattice subjected to a sudden increase of the atom-molecule coupling $g$ from zero.
 
As an initial condition, we assume that the system is deep in a Mott insulating state of atoms with double filling and that the atom-molecule coupling $g$ is zero. In such a situation we can safely neglect the hopping term of atoms as long as we are interested in the atom-loss dynamics and its timescale $1/\Gamma_{\rm PA}$ is much shorter than that of the hopping, which is on the order of $\hbar U/J^2$ in the Mott insulator. Thus, the system can be described by the following single-site master equation,
\begin{eqnarray}
\hbar \frac{d}{dt}\hat{\rho}^{\rm loc} = - i \left[\hat{H}^{\rm loc}, \hat{\rho}^{\rm loc}\right] 
+ L_{\rm M}^{\rm loc}(\hat{\rho}^{\rm loc}).
\label{eq:master_local_AM}
\end{eqnarray}
The Hamiltonian is given by
\begin{eqnarray}
\hat{H}^{\rm loc} = \hat{H}_0^{\rm loc} + \hat{H}_{\rm am}^{\rm loc},
\end{eqnarray}
where
\begin{eqnarray}
\hat{H}_0^{\rm loc} &=& \left( D \hat{n}_{\rm M} + \frac{U}{2} \hat{n}_{\rm A}(\hat{n}_{\rm A}-1) 
+ W \hat{n}_{\rm A}\hat{n}_{\rm M}\right),
\\
\hat{H}_{\rm am}^{\rm loc} &=& g\left( \hat{m}^{\dagger} \hat{a}\hat{a} + {\rm h.c.} \right).
\end{eqnarray}
The one-body molecular loss term is given by
\begin{eqnarray}
L_{\rm M}^{\rm loc}(\hat{\rho}^{\rm loc}) = \frac{\hbar \Gamma_{\rm M}}{2} \left(
- \hat{n}_{\rm M} \hat{\rho}^{\rm loc} - \hat{\rho}^{\rm loc} \hat{n}_{\rm M} + 2\hat{m} \hat{\rho} \hat{m}^{\dagger}\right). 
\end{eqnarray}

Since the initial state is $|n_{\rm A}=2,n_{\rm M}=0\rangle$, the Hilbert space necessary for describing the dynamics of Eq.~(\ref{eq:master_local_AM}) is spanned by only three states, namely, $|n_{\rm A}=0,n_{\rm M}=0\rangle$, $|n_{\rm A}=2,n_{\rm M}=0\rangle$, and $|n_{\rm A}=0,n_{\rm M}=1\rangle$. When $\hbar\Gamma_{\rm M} \gg |g|$, we can properly eliminate the state $|n_{\rm A}=0,n_{\rm M}=1\rangle$ by means of the perturbation theory used in the previous section, to derive the effective master equation,
\begin{eqnarray}
\hbar\frac{d}{dt}\hat{\rho}_{\rm eff}^{\rm loc} = -i \left[ 
\hat{H}_{\rm eff}^{\rm loc}, \hat{\rho}_{\rm eff}^{\rm loc}
\right]
+ L_2^{\rm loc}(\hat{\rho}_{\rm eff}^{\rm loc}),
\label{eq:master_local_BH}
\end{eqnarray}
where
\begin{eqnarray}
\hat{H}_{\rm eff}^{\rm loc} = \frac{U}{2} \hat{n}_{\rm A}(\hat{n}_{\rm A}-1) ,
\end{eqnarray}
\begin{eqnarray}
L_2^{\rm loc}(\hat{\rho}_{\rm eff}^{\rm loc}) = 
\frac{\hbar \tilde{\Gamma}_{\rm PA}}{4} \left(
- \hat{a}^{\dagger}\hat{a}^{\dagger}\hat{a}\hat{a} \hat{\rho}_{\rm eff}^{\rm loc}
- \hat{\rho}_{\rm eff}^{\rm loc} \hat{a}^{\dagger}\hat{a}^{\dagger}\hat{a}\hat{a}
+ 2 \hat{a}\hat{a} \hat{\rho}_{\rm eff}^{\rm loc}\hat{a}^{\dagger}\hat{a}^{\dagger}
\right),
\end{eqnarray}
and 
\begin{eqnarray}
\tilde{\Gamma}_{\rm PA} = 8\frac{g^2}{\hbar^2 \Gamma_{\rm M}}
\left(1 + 4\left(\frac{D-U}{\hbar \Gamma_{\rm M}}\right)^2 \right)^{-1}.
\label{eq:Lorentzian}
\end{eqnarray}
In contrast to the case in the previous section, the effective model is valid without the condition that $\hbar \Gamma_{\rm M}\gg \max(|D|,|U|,|W|)$ and we could obtain the analytical expression of $\tilde{\Gamma}_{\rm PA}$, which includes the explicit dependence on $U$ and $D$. It is obvious that when $\hbar\Gamma_{\rm M} \gg \max(|D|,|U|,|W|)$, $\tilde{\Gamma}_{\rm PA}$ coincides with $\Gamma_{\rm PA}$.

\begin{figure}[tbp]
\includegraphics[scale=0.9]{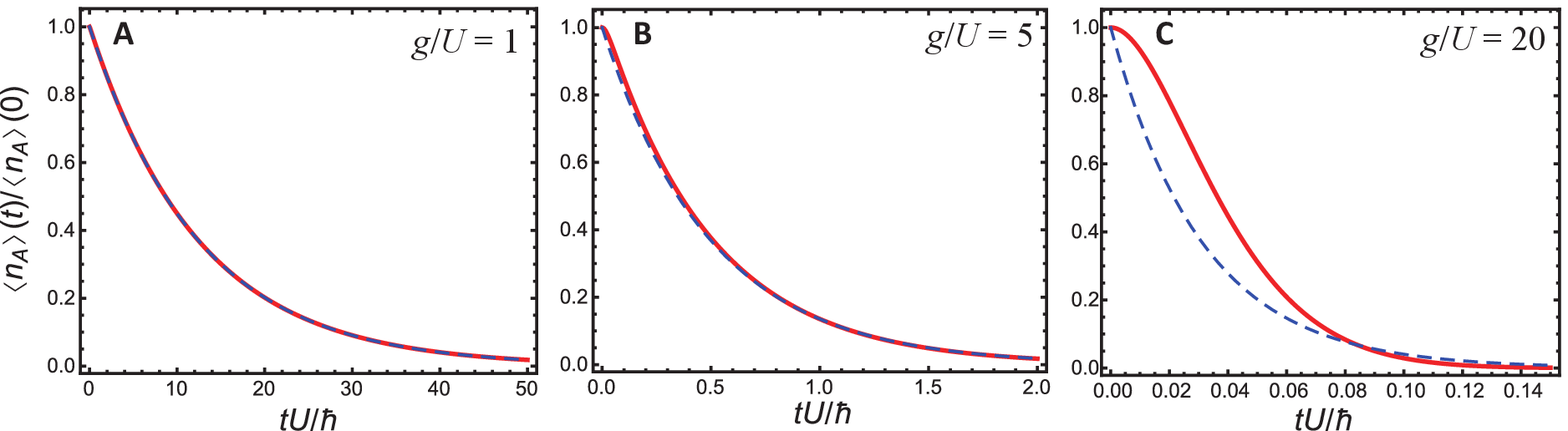}
  \caption{
\textbf{Time evolution of the normalized atom density $\langle \hat{n}_{\rm A} \rangle(t)$ for $\langle \hat{n}_{\rm A} \rangle(0) = 2$.} 
We take $\hbar\Gamma_{\rm M}/U =100$ and $D/U=1$. 
The red solid lines represent the numerical solution of Eq.~(\ref{eq:master_local_AM}) while the blue dashed line represent the analytical solution (\ref{eq:nAoft}) of the effective model (\ref{eq:master_local_BH}).
  }
\label{fig:aLoss}
\end{figure}

The dynamics of the effective master equation (\ref{eq:master_local_BH}) involves only the two states such that we can easily obtain its analytical solution,
\begin{eqnarray}
\langle\hat{n}_{\rm A}\rangle(t) = 2 e^{-\tilde{\Gamma}_{\rm PA}t}.
\label{eq:nAoft}
\end{eqnarray}
In order to check the validity of the effective master equation (\ref{eq:master_local_BH}),  in Fig.~\ref{fig:aLoss} we compare Eq.~(\ref{eq:nAoft}) with the numerical solution of the original master equation (\ref{eq:master_local_AM}) including the molecular degree of freedom. We see that the analytical and numerical results agree when $\hbar\Gamma_{\rm M}\gg g$.

\begin{figure}[tbp]
\includegraphics[scale=1.2]{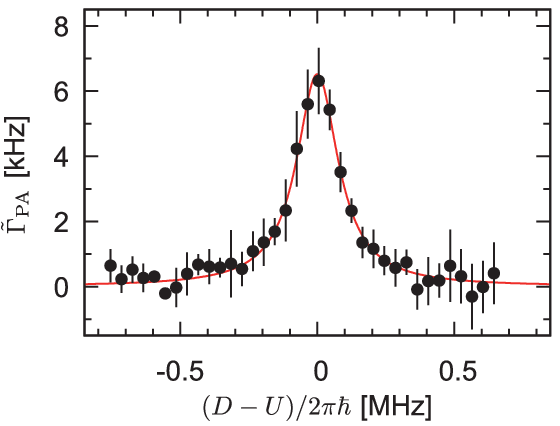}
  \caption{
\textbf{Measurement of the one-body molecular loss $\tilde{\Gamma}_{\rm PA}$.} 
The black points represent the loss rate while the red curve is the fit of Eq.~(\ref{eq:Lorentzian}).
}
\label{fig:PAlinewidth}
\end{figure}

The above results indicate that when the condition $\hbar \Gamma_{\rm M} \gg g$ is safely satisfied, one can determine $\tilde{\Gamma}_{\rm PA}$ in experiment by measuring the atom-loss dynamics in the double-filling Mott insulating state and extracting the exponent in the exponential decay of the atom number. Once $\tilde{\Gamma}_{\rm PA}$ is measured as a function of the detuning $D$, $\Gamma_{\rm M}$ can be also determined by fitting $\tilde{\Gamma}_{\rm PA}(D)$ to the Lorentzian function of Eq.~(\ref{eq:Lorentzian}).
In Fig. \ref{fig:PAlinewidth}, we show measured $\tilde{\Gamma}_{\rm PA}$ as a function of $D-U$. Fitting the data to Eq.~(\ref{eq:Lorentzian}), we determine $\Gamma_{\rm M} = 2\pi \times $185(13) kHz = 1.16(8) MHz.
On the other hand, $g$ can be estimated from the experimentally determined $\Gamma_{\rm M}$ and $\Gamma_{\rm PA}$, resulting in $g/\hbar \sim 100~{\rm kHz}$ at most.
Therefore we can confirm that the condition $\hbar\Gamma_{\rm M} \gg g$ is satisfied, as well as the condition $\hbar\Gamma_{\rm M} \gg  \max(|D|,|U|,|W|)$ since $D/\hbar$, $U/\hbar$ and $W/ \hbar$ are at most a few 10 kHz in our experiment.

\section{Details of the theoretical analyses using the Gutzwiller variational approach}
In this section, we present some details of theoretical calculations regarding dynamics starting from the Mott insulator with one bosonic atom per site in the presence of the atom-molecule coupling that leads to the loss of atoms. The corresponding dynamics is experimentally analyzed in the main text. For this purpose, we first review how to solve the master equations (\ref{eq:master_init}) and (\ref{eq:master_BH}) within the Gutzwiller mean-field approximation~
\cite{diehl-10,tomadin-11}. 
Using the introduced prescription, we next compute the atom-loss dynamics after a fast ramp-down of the lattice depth and that  during a slow ramp-down of the lattice depth.

Here we explicitly explain the Gutzwiller mean-field theory applied to the atom-molecule mixture model of Eq.~(\ref{eq:master_init}) because it contains the molecular degrees of freedom, which has not been taken into account in previous studies. We note that one can easily apply the same prescription to the effective Bose-Hubbard model of Eq.~(\ref{eq:master_BH}) in a very similar manner.
In the Gutzwiller mean-field approximation, the many-body density matrix is assumed to be a single product of local density matrices,

\begin{eqnarray}
\hat{\rho} = \prod_j \hat{\rho}_j^{\rm GW},
\end{eqnarray}
where
\begin{eqnarray}
\hat{\rho}_j^{\rm GW} = \sum_{l_j = 1}^{2d} \sum_{m_j = 1}^{2d}\rho_{l_j,m_j}^{(j)}|l_j\rangle \langle m_j|_j.
\end{eqnarray}
From the Gutzwiller density matrix, we define the local superfluid order parameter as
\begin{eqnarray}
\psi_{j} = \langle \hat{a}_j \rangle = {\rm Tr}[\hat{\rho}_j^{\rm GW}\hat{a}_j].
\end{eqnarray}
In the Gutzwiller mean-field approximation, we ignore the second order terms with respect to the fluctuation of $\hat{a}_j$ from its mean value $\psi_j$. In this way, the Hamiltonian is simplified as
\begin{eqnarray}
\hat{H} \simeq \sum_{j} \hat{H}_j^{\rm GW},
\end{eqnarray}
where
\begin{eqnarray}
\hat{H}_j^{\rm GW} = \hat{H}_{0,j}^{\rm loc} 
+ \hat{H}_{{\rm am},j}^{\rm loc} 
+ \hat{H}_{{\rm hop},j}^{\rm GW},
\end{eqnarray}
\begin{eqnarray}
\hat{H}_{{\rm hop},j}^{\rm GW} = - J \sum_{\langle k \rangle_j} 
\left( \psi_k^{\ast}\hat{a}_j + \hat{a}_j^{\dagger}\psi_k \right).
\label{eq:hop_GW}
\end{eqnarray}
In Eq.~(\ref{eq:hop_GW}), $\langle k \rangle_j$ means sites nearest-neighboring to $j$. It is worth noting that $\hat{H}_j^{\rm GW}$ includes only local operators at site $j$ because one of field operators in the hopping term is replaced with its mean value. Thanks to this property, the master equation under the Gutzwiller mean-field approximation is closed within local site $j$,
\begin{eqnarray}
\hbar\frac{d}{dt}\hat{\rho}^{\rm GW}_j = - i \left[\hat{H}^{\rm GW}_j, \hat{\rho}^{\rm GW}_j\right] 
+ L_{\rm M}^{\rm loc}(\hat{\rho}^{\rm GW}_j),
\label{eq:master_GW}
\end{eqnarray}
such that we can solve the master equation at a very low numerical cost.

While the Gutzwiller approximation is a simple mean-field theory, it has been extensively used to study various phenomena and properties of Bose gases in optical lattices, including the quantum phase transitions~\cite{rokhsar-91,sheshadri-93,iskin-11}, 
the elementary excitations~\cite{kovrizhin-05, krutitsky-11}, 
the superfluid critical momentum~\cite{altman-05,saito-12}, 
and the non-equilibrium dynamics~\cite{snoek-07,bissbort-11,Snoek-11}. 
Recently, it has been applied for solving the master equation of the Bose-Hubbard system with dissipation terms~\cite{diehl-10, tomadin-11, leboite-13, vidanovic-14}. 
This approximation is more accurate in higher dimensions, where there are more mean fields to interact with. In the case of the Bose-Hubbard model on a cubic lattice, for instance, the Gutzwiller approximation gives the critical point for the superfluid-Mott insulator quantum phase transition at unit filling as $zJ/U=0.1716$ while that by the quantum Monte Carlo method is $zJ/U=0.2045$~\cite{capogrosso-07}, 
where $z$ is the coordination number. Since the experimental system considered here is three dimensional, the Gutzwiller approximation can give reliable results at least qualitatively.   

Assuming that the system is homogeneous, we further simplify the master equation, i.e., we drop the dependence on the site index $j$. This assumption means that we neglect fluctuations other than the zero-momentum one and the effect of the trapping potential. Since the trapping potential is present in the actual experiment, our theoretical analyses within this simplification do not correspond to the experiment at a quantitative level. We emphasize that the main purpose of our theoretical analyses is to provide qualitative explanations for the interesting effects of the engineered dissipation observed in the experiment.

\subsection{Dynamics after a fast ramp-down of the lattice depth}
We consider the atom-loss dynamics from the initial unit-filling Mott insulator induced by the PA laser after a fast ramp-down of the lattice depth. The corresponding dynamics is experimentally analyzed in the main text (see, e.g., Fig.~2). The optical lattice potential is given by
\begin{eqnarray}
V_{\rm ol}({\boldsymbol r}) = V_0 \left(\sin^2(kx) + \sin^2(ky) + \sin^2(kz)\right),
\end{eqnarray}
where $k=\pi/d$ with the lattice constant $d=266\,{\rm nm}$.
We set the time sequence of the lattice depth,
\begin{eqnarray}
V_0(t) = \left\{
\begin{array}{cc}
v_{\rm up} (t - t_{0}) + V_{0,{\rm ini}}, & {\rm when}\,\, t_{0}\leq t < t_{1},
\\
v_{\rm down} (t - t_{1}) + V_{0,{\rm max}}, & {\rm when}\,\, t_{1} \leq t < t_{2},
\\
V_{0,{\rm fin}}, & {\rm when}\,\, t_{2}\leq t < t_{\rm fin},
\end{array}
\right.
\label{eq:Vsequence1}
\end{eqnarray}
where the ramp-up and ramp-down speeds are given by
\begin{eqnarray}
v_{\rm up} = \frac{V_{0,{\rm max}}-V_{0,{\rm ini}}}{t_{1}-t_{0}},
\label{eq:vup}
\\
v_{\rm down} = \frac{V_{0,{\rm fin}}-V_{0,{\rm max}}}{t_{2}-t_{1}}.
\label{eq:vdown}
\end{eqnarray}

\begin{figure}[tbp]
\includegraphics[scale=1.0]{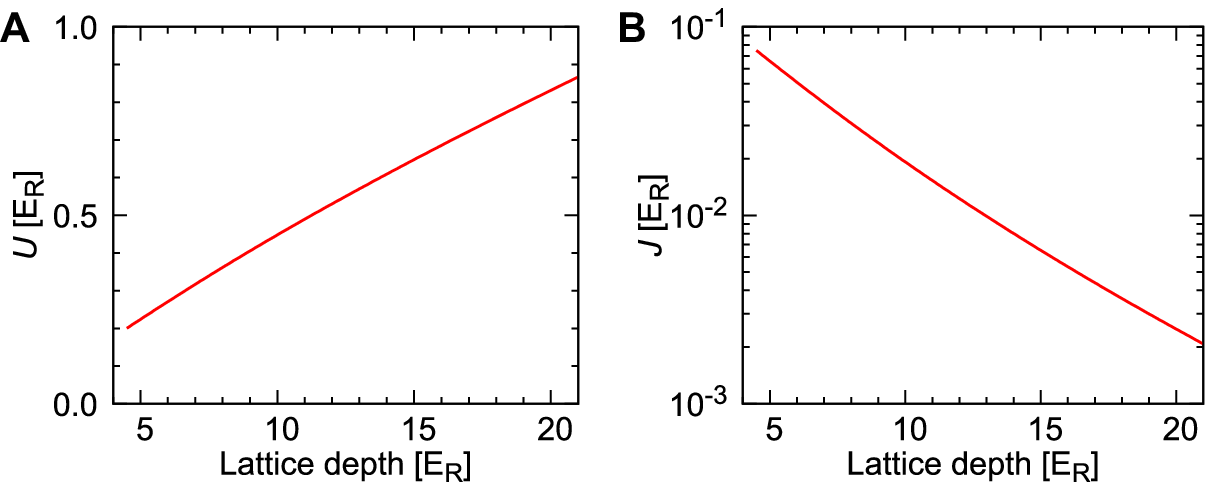}
  \caption{
\textbf{The on-site interaction $U$ and the hopping energy $J$ as a function of the lattice depth.} 
\textbf{(A)}, The on-site interaction $U$. \textbf{(B)}, The hopping energy $J$.
}
\label{fig:UJvsLatticeDepth}
\end{figure}

Once the lattice depth is given, we numerically compute the Wannier function $w_j({\boldsymbol r})$ localized at site $j$, from which we can determine the on-site interaction and the hopping energy as
\begin{eqnarray}
U &=& \frac{4\pi \hbar^2 a_{s}}{m} \int d{\boldsymbol r} |w_j({\boldsymbol r})|^4, \\
J &=& - \int  d{\boldsymbol r} w_j^{\ast}({\boldsymbol r})\left(- \frac{\hbar^2\nabla^2}{2m} 
+ V_{\rm ol}({\boldsymbol r}) \right)w_k^{\ast}({\boldsymbol r}).
\end{eqnarray}
Sites $j$ and $k$ are assumed to be nearest neighboring. The $s$-wave scattering length of $^{174}$Yb used in the experiment is $a_s = 5.55 \, {\rm nm}$~\cite{kitagawa-08}. 
In Fig.~\ref{fig:UJvsLatticeDepth}, we plot the on-site interaction $U$ and the hopping energy $J$ as a function of the lattice depth.

\begin{figure}[tbp]
\includegraphics[scale=0.8]{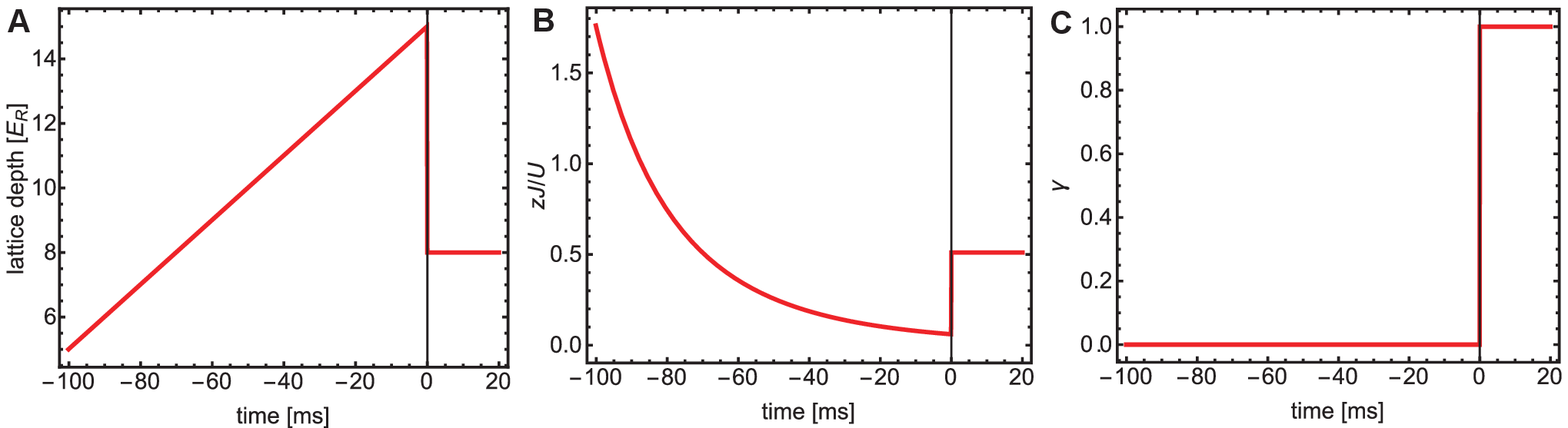}
  \caption{
\textbf{Time sequence of the atom-loss measurement from the Mott insulating state with unit filling.} 
\textbf{(A)}, The lattice depth, which is expressed in Eq.~(\ref{eq:Vsequence1}). \textbf{(B)}, $zJ/U$. \textbf{(C)}, $\gamma =\hbar\Gamma_{\rm PA}/U$ in the case that $\gamma(t>0)=1$.
}
\label{fig:sequence1}
\end{figure}

Let us elaborate the time sequence of the dynamics computed here. We start with the superfluid ground-state at $V_{0,{\rm ini}} = 5 E_{\rm R}$ and $t_0 = -100.2\,{\rm ms}$. Strictly speaking, in the experiment, there is an additional ramp-up process from $V_{0} = 0 E_{\rm R}$ to $5 E_{\rm R}$, but we can not take into account this process because the Bose-Hubbard model is invalid for such a shallow optical lattice. Nevertheless, the ramp-up speed in the experiment is so slow that the superfluid state prepared at $V_{0,{\rm ini}} = 5 E_{\rm R}$ can be regarded as the ground state. We next prepare the Mott insulating state at $V_{0,{\rm max}} = 15 E_{\rm R}$ and $t_1 = -0.2\,{\rm ms}$, which imply that the ramp-up speed is $v_{\rm up} = 0.1\,E_{\rm R}/{\rm ms}$. Right after the preparation of the Mott insulating state, we ramp down the lattice depth to $V_{0,{\rm fin}}=8 E_{\rm R}$ in $0.2\,{\rm ms}$, implying that $v_{\rm down} = -35\,E_{\rm R}/{\rm ms}$ and $t_{\rm 2}=0$. Finally, we turn on the dissipation term $\Gamma_{\rm PA}$ and keep $V_0(t)=V_{0,{\rm fin}}$ until $t_{\rm fin} = 20 \,{\rm ms}$. The time sequence for $V_0/E_{\rm R}$, $zJ/U$, and $\gamma = \hbar\Gamma_{\rm PA}/U$ is summarized in Fig.~\ref{fig:sequence1}.

\begin{figure}[tbp]
\includegraphics[scale=0.7]{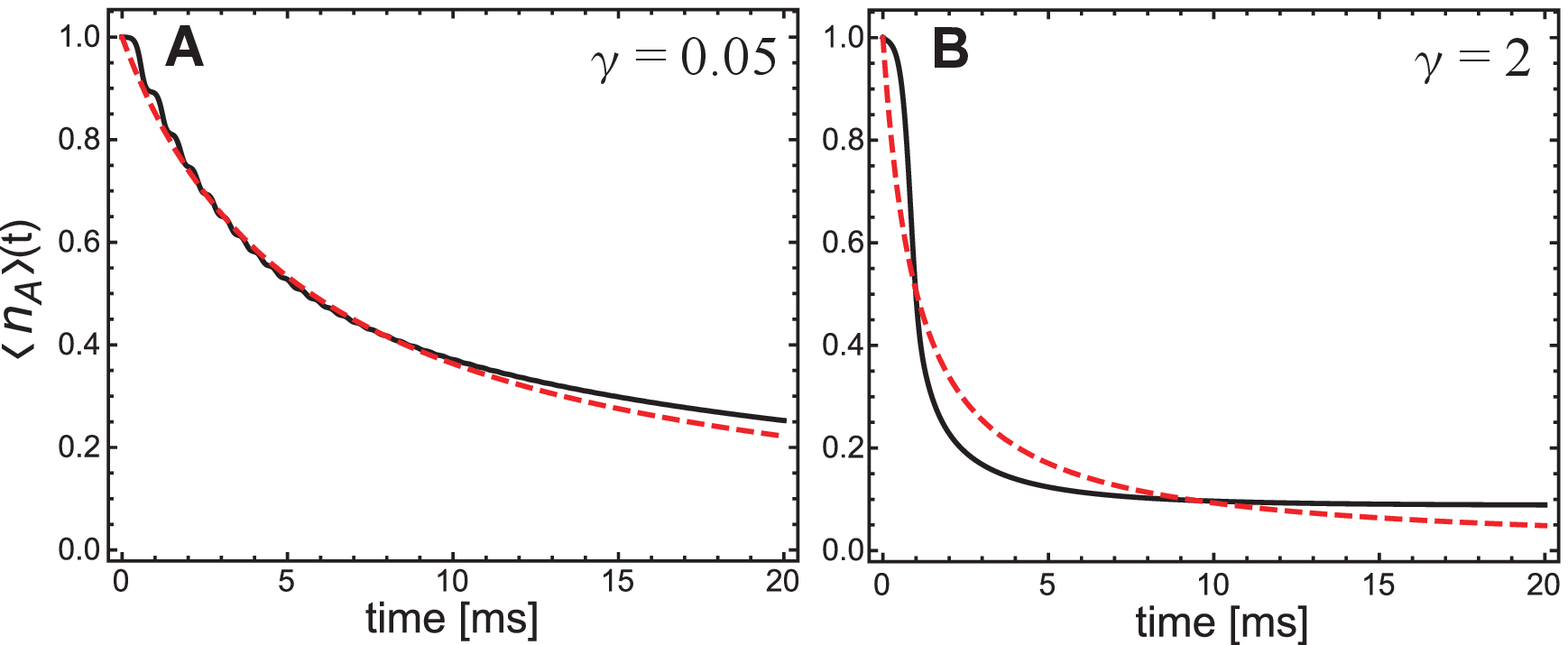}
  \caption{
\textbf{Time evolution of the atom density $\langle \hat{n}_{\rm A}\rangle(t)$ for $\langle \hat{n}_{\rm A} \rangle(0) = 1$. }
The black solid curve represents the numerical solution of the effective master equation (\ref{eq:master_BH}). The red dashed line represents the fitting through the function of Eq.~(\ref{eq:fitting1}). The fitting is made for the data satisfying the condition that $\langle \hat{n}_{\rm A}\rangle(t)>0.4$.  
  }
\label{fig:aLoss_unit}
\end{figure}

\begin{figure}[tbp]
\includegraphics[scale=0.5]{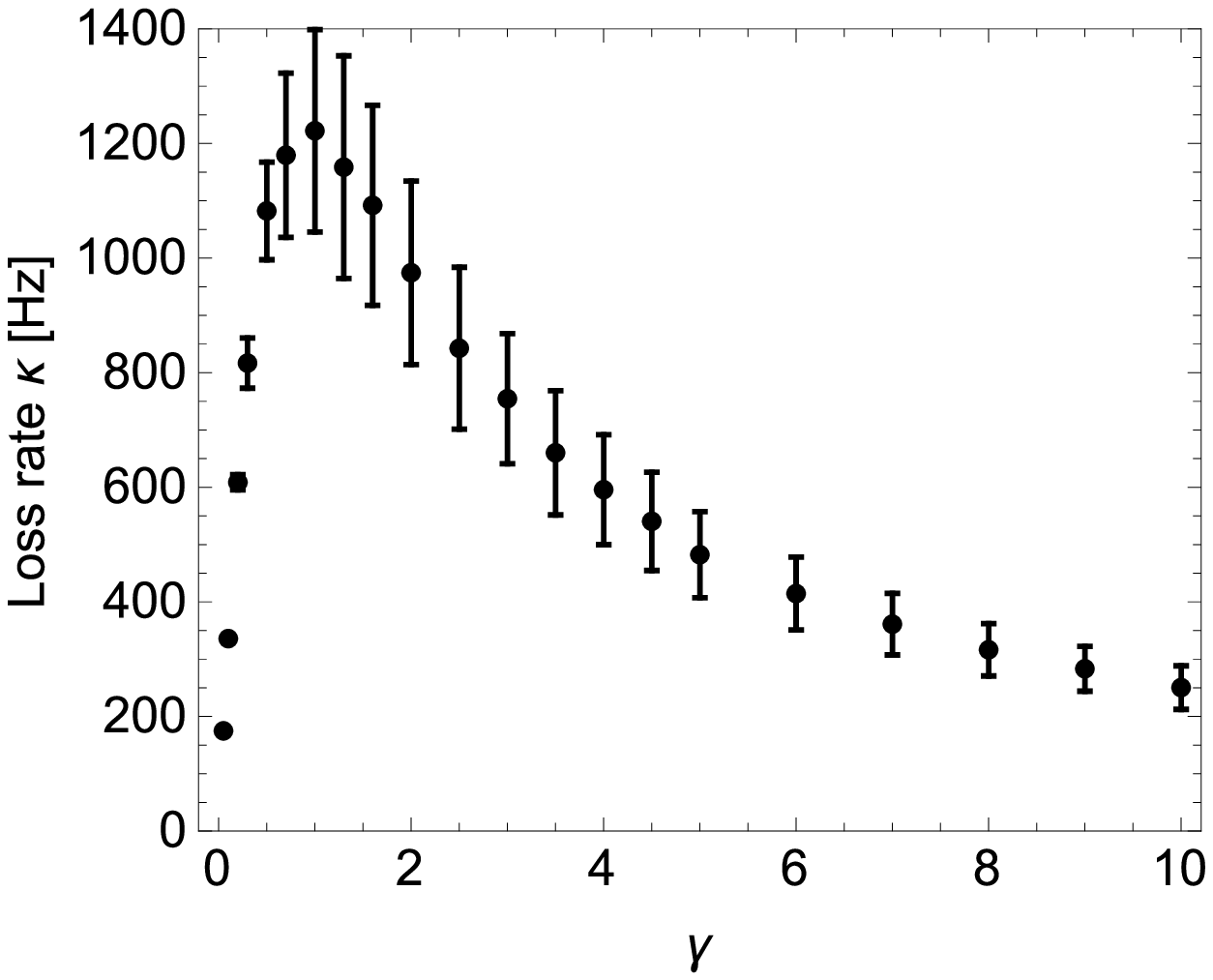}
  \caption{
\textbf{Loss rate $\kappa$ as a function of the dissipation strength $\gamma$.}  
$\kappa$ is extracted by the fitting to the time evolution of the atom density $\langle \hat{n}_{\rm A}\rangle(t)$ starting from the Mott insulating state with unit filling. 
  }
\label{fig:lossVsG}
\end{figure}

In Fig.~\ref{fig:aLoss_unit}, we plot the time evolution of the atom number per site $\langle \hat{n}_{\rm A} \rangle$ at $t>t_2=0$, where $\gamma = 0.05$ (A) and $2$ (B). 
We obviously see that $\langle \hat{n}_{\rm A} \rangle$ decays as the time evolves. In order to extract the loss rate from the numerical data, we use the fitting function,
\begin{eqnarray}
f(t) = \frac{1}{1+\kappa t},
\label{eq:fitting1}
\end{eqnarray}
where the loss rate $\kappa$ is treated as a free parameter. Notice that the same fitting function is used to extract the loss rate from the experimental data (see Fig.~1 (C) of the main text). In Fig.~\ref{fig:lossVsG}, the extracted loss rate is plotted as a function of $\gamma$. While the function of Eq.~(\ref{eq:fitting1}) is well fitted to $\langle \hat{n}_{\rm A} \rangle(t)$ at small $\gamma$ as shown in Fig.~\ref{fig:aLoss_unit}(A), it is worse for relatively large $\gamma$ (see, e.g., Fig.~\ref{fig:aLoss_unit}(B)). This leads to small and large error bars for small and large $\gamma$, respectively.

\begin{figure}[tbp]
\includegraphics[scale=0.7]{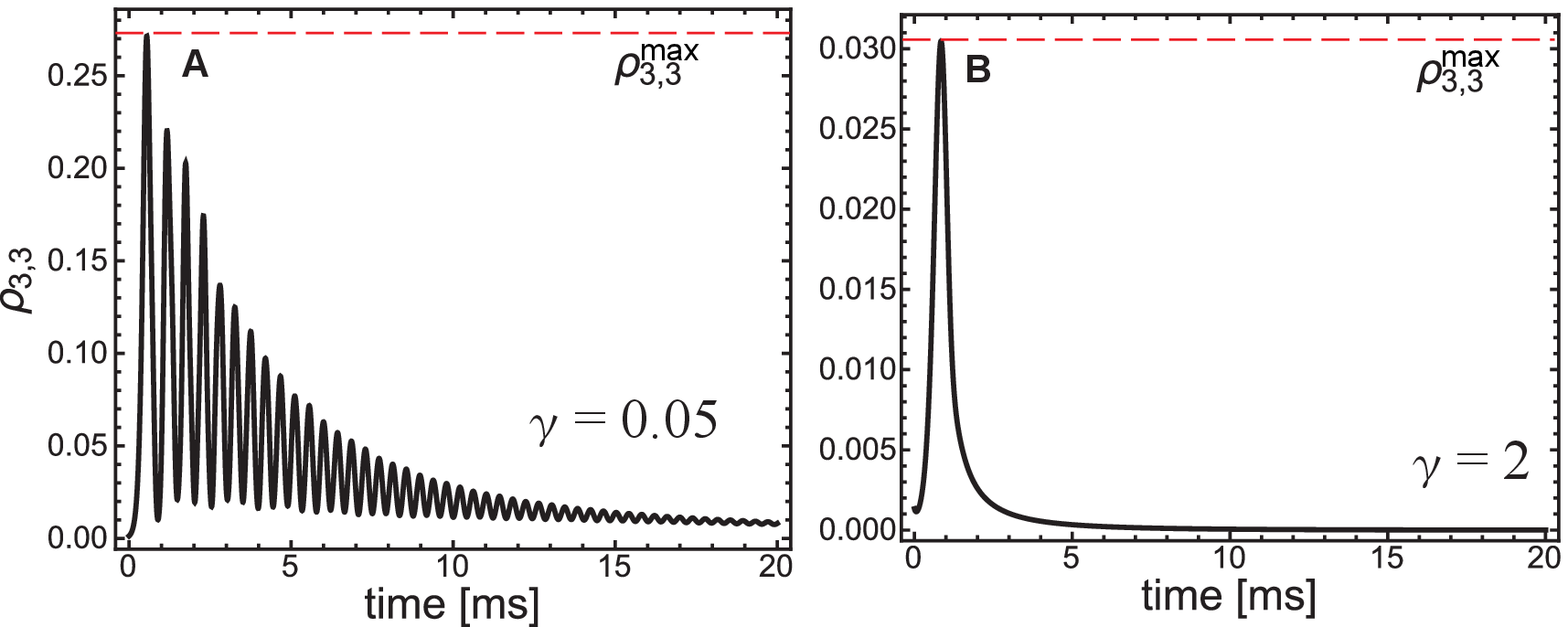}
  \caption{
\textbf{Time evolution of $\rho_{3,3}$.} 
The red dashed line indicates its maximum value $\rho_{3,3}^{\rm max}$.
}
\label{fig:rho33}
\end{figure}

\begin{figure}[tbp]
\includegraphics[scale=0.7]{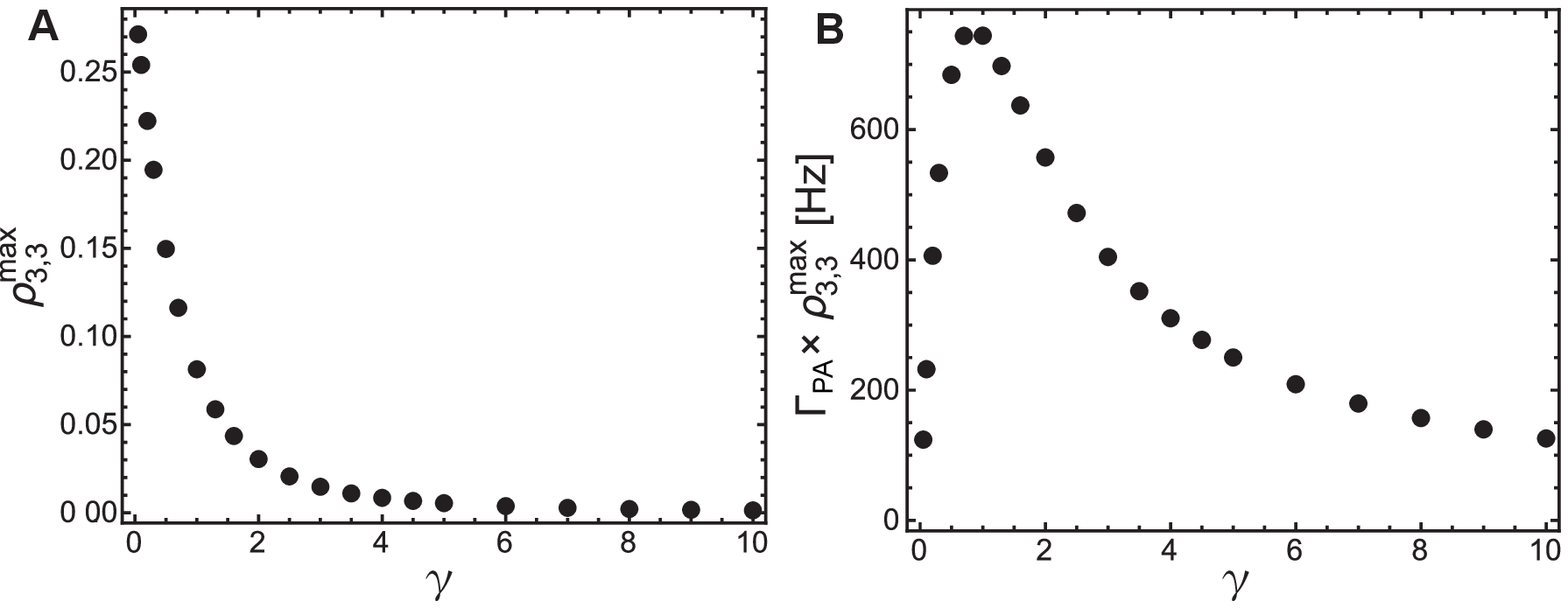}
  \caption{
\textbf{$\rho_{3,3}^{\rm max}$ and $\Gamma_{\rm PA} \times \rho_{3,3}^{\rm max}$ as a function of $\gamma$.} 
}
\label{fig:rho33max}
\end{figure}

We see from Fig.~\ref{fig:lossVsG} that when the dissipation strength $\gamma$ increases, the loss rate $\kappa$ initially increases but starts to decrease around $\gamma=1$. A similar behavior is also seen in the experiment as shown in Fig.~1 (C) of the main text. The decreasing loss rate at $\gamma \gtrsim 1$ can be attributed to the suppression of the double occupancy due to the strong two-body loss, namely the continuous Zeno effect. In order to corroborate this interpretation, we show the time evolution of $\rho_{3,3}$, which is an element of the local density matrix corresponding to the probability of the double occupancy, in Fig.~\ref{fig:rho33}. We take the maximum value of $\rho_{3,3}$ during the time evolution at $t>t_2=0$ and plot it as a function of $\gamma$ in Fig.~\ref{fig:rho33max}(A). We clearly see that the double occupancy is strongly suppressed due to the strong dissipation. Roughly speaking, the loss rate can be estimated by $\Gamma_{\rm PA} \times \rho_{3,3}^{\rm max}$, which is plotted as a function of $\gamma$ in Fig.~\ref{fig:rho33max}(B). There we see the tendency similar to the actual loss rate shown in Fig.~\ref{fig:lossVsG}.

\subsection{Dynamics during a slow ramp-down of the lattice depth}
We consider the dynamics of the initial unit-filling Mott insulator during a slow ramp-down of the lattice depth towards the supefluid state. 
By analyzing this type of dynamics, we aim to understand the effect of the two-body loss term $\Gamma_{\rm PA}$ on the quantum phase transition between the superfluid and Mott insulator. 
As mentioned in the main text, an important effect is that the two-body loss term explicitly breaks the conservation of the particle number of the system. Since the superfluid-Mott insulator transition at $\Gamma_{\rm PA}=0$ is originated from the U(1) symmetry associated with the particle-number conservation, the introduction of finite $\Gamma_{\rm PA}$ changes the transition to a crossover. Notice, however, that the explicit breaking of the particle-number conservation does not mean that of the U(1) symmetry in our dissipative system. This is in clear contrast to a closed system with no dissipation.

Another important effect is that the two-body loss term makes the ``superfluid" state at unit filling so dissipative that it cannot carry dissipationless superflow. In this sense, even a small loss term immediately breaks the superfluidity. 
However, one can distinguish such a lossy gas with delocalized atoms and long-range coherence from the Mott insulating state and study the crossover from the latter state to the former in the dynamics subjected to a slow ramp-down of the lattice depth. 
Hereafter, for convenience we use the term ``superfluid" to describe the former state.

Before recreating the experimental situation, we present some important properties of the crossover phenomenon that are independent of either the preparation procedure of the initial Mott insulating state or the ramp-down speed within the Gutzwiller approximation. As elaborated below, we specifically focus on the growth rate of the superfluid order parameter amplitude during a slow increase of the hopping energy or the on-site interaction.
The time sequence of the hopping energy is given by
\begin{eqnarray}
J(t) = \frac{J_{\rm fin} - J_{\rm ini}}{\tau}t + J_{\rm ini},
\end{eqnarray}
while the other parameters are fixed to be time-independent, where $\tau$ denotes the total evolution time. In the case that we vary the on-site interaction, its time sequence is given  by
\begin{eqnarray}
U(t) = \frac{U_{\rm fin} - U_{\rm ini}}{\tau}t + U_{\rm ini}.
\end{eqnarray}
In our experiment shown in the main text, we ramp down the lattice depth to cause the crossover from the Mott insulator to the ``superfluid". Since the dominant effect of the ramp down of the lattice depth is the exponential increase of the hopping energy, the case of $J$ increase is closer to the experimental situation. Notice that our ${}^{174}$Yb atoms in their electronic ground state do not have usable Feshbach resonance such that we can not dynamically control $U$ to realize the case of $U$ decrease.

We set the initial and final values of the hopping energy as $zJ_{\rm ini}/U=0$ and $zJ_{\rm fin}/U=0.5$, and those of the on-site interaction energy as $U_{\rm ini}/(zJ)=50$ and $U_{\rm fin}/(zJ)=0.5$. At $zJ/U=zJ_{\rm ini}/U$ or $zJ/U_{\rm ini}$ at unit filling, the ground state is the Mott insulating state, i.e., $\rho_{l,m}^{\rm gs}=\delta_{l,2}\delta_{m,2}$. As an initial state of the dynamics, we add small random noise terms to $\rho_{l,m}^{\rm gs}$ as
\begin{eqnarray}
\rho_{l,m}(t=0) = \rho_{l,m}^{\rm gs} + \epsilon_{l,m}^{\rm re} + i \epsilon_{l,m}^{\rm im},
\end{eqnarray}
where $\epsilon_{l,m}^{\rm re}$ and $\epsilon_{l,m}^{\rm im}$ are assumed to be independent random variables with zero average and a box distribution from $-\varepsilon$ to $\varepsilon$.
In the absence of the noise terms ($\varepsilon=0$), the system remains in the initial state and the evolution towards the ``superfluid" state can not be captured because  $\rho_{l,m}^{\rm gs}$ is a time-independent solution of the effective master equation (\ref{eq:master_BH}) within the Gutzwiller approximation. 

\begin{figure}[tbp]
\includegraphics[scale=0.8]{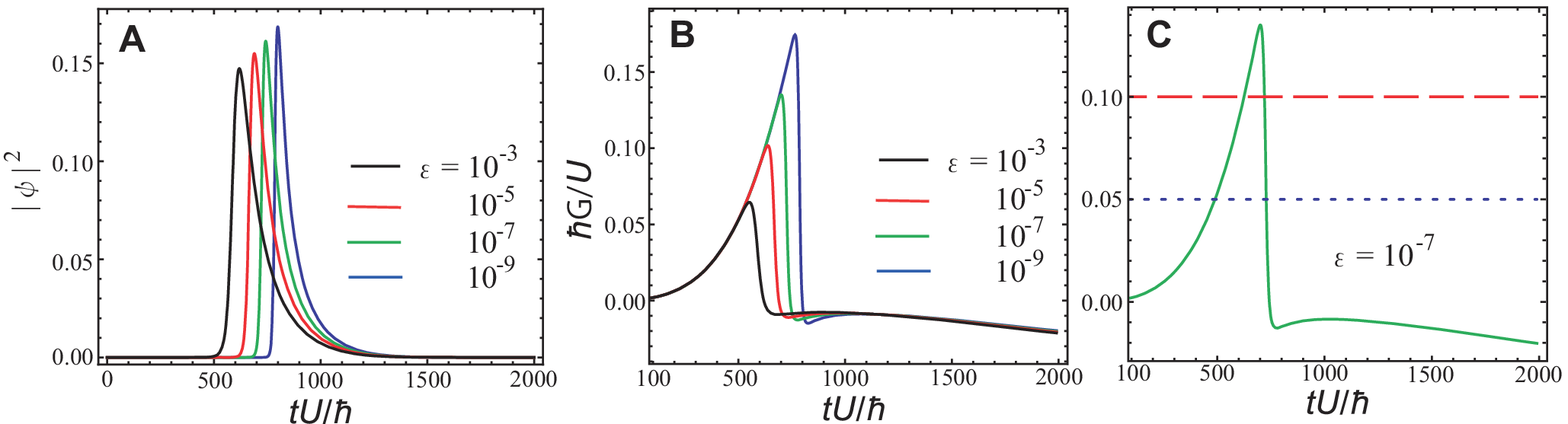}
  \caption{
\textbf{Time evolution of the amplitude of the superfluid order parameter and its growth rate.} 
\textbf{(A)}, Time evolution of $|\psi|^2$. \textbf{(B)}, \textbf{(C)}, Growth rate $G = \frac{d}{dt}\ln{|\psi|^2}$ during the linear ramp-up of the hopping $J$, where $\tau U/\hbar = 2000 $, $zJ_{\rm ini}/U = 0.0$, $zJ_{\rm fin}/U=0.5$, and $\gamma = 1$.
  }
\label{fig:psi}
\end{figure}

\begin{figure}[tbp]
\includegraphics[scale=0.5]{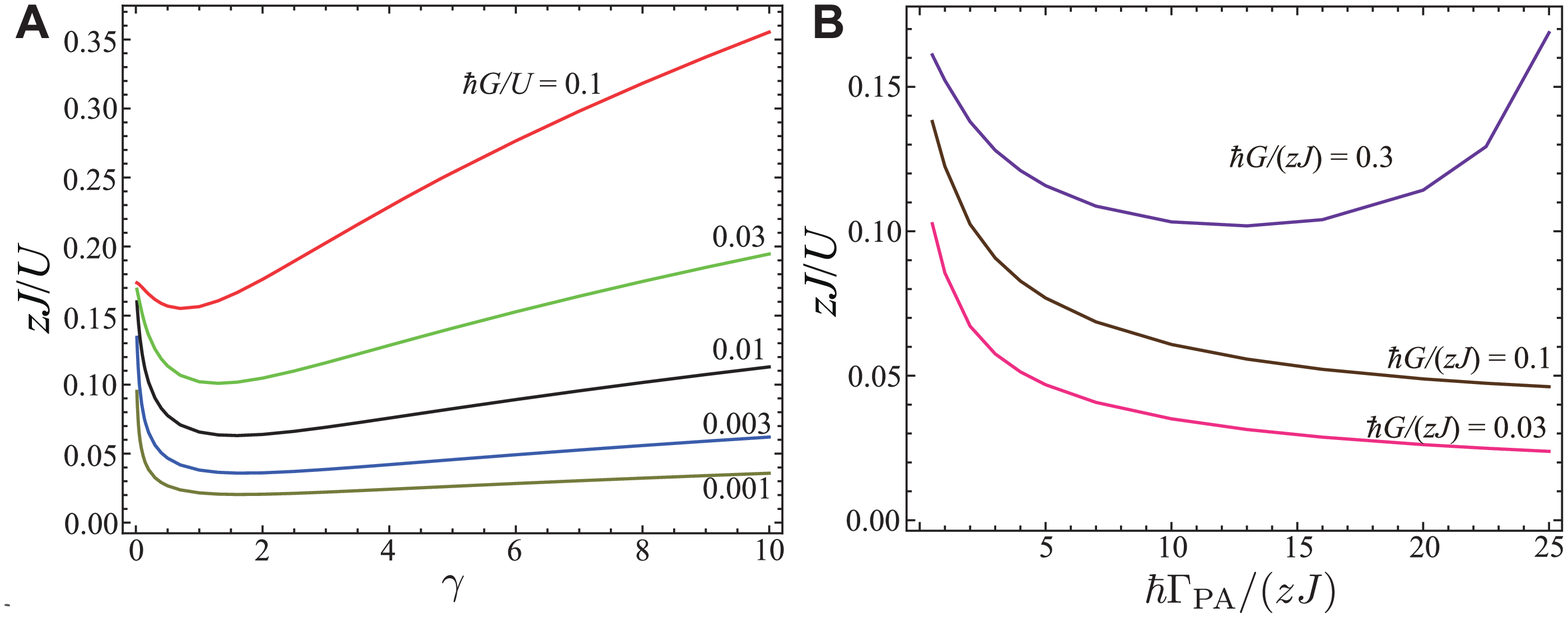}
  \caption{
\textbf{Contour plot of $G=\frac{d}{dt}\ln{|\psi|^2}$ in the $(\gamma,zJ/U)$-plane (A) and that in the $(\hbar\Gamma_{\rm PA}/(zJ),zJ/U)$-plane (B).} 
While we take $zJ_{\rm ini}/U = 0.0$, $zJ_{\rm fin}/U=0.5$, and $\tau U/\hbar = 10000$ for (A) and $U_{\rm ini}/(zJ)=50$, $U_{\rm fin}/(zJ)=0.5$, and $\tau zJ/\hbar = 2000$ for (B), the contour plot is independent of these parameters as long as $\tau$ is sufficiently large.
}
\label{fig:CrossOver}
\end{figure}

In Fig.~\ref{fig:psi}(A), we show the time evolution of the amplitude of the superfluid order parameter $|\psi|^2$ for several values of the noise strength $\varepsilon$. We see that $|\psi|^2$ significantly depends on $\varepsilon$. In contrast, as shown in Fig.~\ref{fig:psi}(B), we find in the time evolution of the rate of the exponential growth in $|\psi|^2$, namely $G=\frac{d}{dt}\ln |\psi|^2$, that there is a time region where $G$ increases and is independent of $\varepsilon$. Notice that $G$ in such a time region is also independent of $\tau$ as long as $\tau$ is sufficiently large. We use the values of $G$ in the time region to characterize the time scales of the crossover from the Mott insulator to the ``superfluid" that are independent of either $\varepsilon$ or $\tau$. As indicated in Fig.~\ref{fig:psi}(C), for given $\gamma$ we determine the value of $zJ/U$ at which $G$ in the universal time region takes a certain value, e.g., $\hbar G/U=0.1$ (red dashed line) or $0.05$ (blue dotted line). In Fig.~\ref{fig:CrossOver}(A), we show a contour plot of $\hbar G/U$ in the $(\gamma,zJ/U)$-plane. There we see that when $\gamma$ increases from zero, the contour lines become more distant from one another, i.e., the transition is changed to a crossover. A similar behavior is also seen in Fig.~\ref{fig:CrossOver}(B), where a contour plot of $\hbar G/(zJ)$ in the $(\hbar \Gamma_{\rm PA}/(zJ),zJ/U)$-plane is shown.
We also see that $zJ/U$ on each contour line exhibits a non-monotonic behavior as a function of $\gamma$; when $\gamma$ increases, it initially decreases but starts to increase above a certain $\gamma$. This result indicates that the strong two-body loss term, i.e., $\gamma \gg 1$, favors the Mott insulating state over the ``superfluid".

We argue that this interesting effect of the two-body loss term on the crossover from the Mott insulator to the ``superfluid", namely the suppression of the dynamical melting of the Mott insulating state, can be observed through the measurement of the atom number or the momentum distribution during the slow ramp-down of the lattice depth. In order to corroborate this argument, we compute the dynamics associated with the change of the lattice depth in time, 
\begin{eqnarray}
V_0(t) = \left\{
\begin{array}{cc}
v_{\rm up} (t - t_{0}) + V_{0,{\rm ini}}, & {\rm when}\,\, t_{0}\leq t < t_{1},
\\
v_{\rm down} (t - t_{1}) + V_{0,{\rm max}}, & {\rm when}\,\, t_{1} \leq t < t_{\rm 2},
\end{array}
\right.
\label{eq:Vsequence2}
\end{eqnarray}
where the ramp-up and ramp-down speeds $v_{\rm up}$ and $v_{\rm down}$ are given by Eqs.~(\ref{eq:vup}) and (\ref{eq:vdown}). We start with the superfluid ground state at $V_{0,{\rm ini}} = 5 \, E_{\rm R}$ and $t_0=-100\,{\rm ms}$ while setting $\gamma = 0$. We slowly ramp up the optical lattice in $100 \,{\rm ms}$ to $V_{0,{\rm max}}=20 \, E_{\rm R}$, implying that $t_1 = 0 \,{\rm ms}$ and $v_{\rm up}=0.15 \, E_{\rm R}/{\rm ms}$, in order to prepare a Mott insulating state. Right after preparing the Mott insulating state, we turn on $\gamma$ to be a finite value and ramp down the optical lattice to $V_{0,{\rm fin}}=5 \, E_{\rm R}$ in $7.5 \,{\rm ms}$, implying that $t_2 = 7.5 \,{\rm ms}$ and $v_{\rm down} = -2 \, E_{\rm R}/{\rm ms}$. Notice that in contrast to the dynamics subjected to the hopping ramp-up analyzed above, we do not explicitly include small random noise terms in the initial condition. Instead, the finite-time ramp-up process creates small excitations in the prepared Mott insulating state at $t=t_1$, which practically take a role of small initial noise terms needed for dynamically melting the initial Mott state into the ``superfluid" state.

In Fig.~\ref{fig:nOfV}, we show the atom number per site during the ramp-down of the lattice depth, where $t_1\le t < t_2$, as a function of the instantaneous value of $V_0/E_{\rm R}$. In Fig.~\ref{fig:nOfV}(A), we see that the onset of the atom loss shifts to the side of large $V_0/E_{\rm R}$ when $\gamma$ increases up to $\gamma=0.5$. In contrast, as shown in Fig.~\ref{fig:nOfV}(B), the onset significantly shifts to the side of small $V_0/E_{\rm R}$ when $\gamma$ increases further from $\gamma=0.5$. This means that the melting of the initial Mott insulating state is delayed due to the effect of the strong two-body loss term. As shown in Fig.~\ref{fig:ncOfV}, a similar tendency is also seen in the dynamics of the condensate fraction $|\psi|^2/\langle \hat{n}_{\rm A} \rangle$, which qualitatively corresponds to the strength of the coherence peak in the momentum distribution. When $\gamma$ increases from $\gamma=0.5$, the onset of the growth of $|\psi|^2/\langle \hat{n}_{\rm A} \rangle$ shifts significantly to the side of small $V_0/E_{\rm R}$. 
Note that the oscillation of the condensate fraction originates from non-adiabaticity of the ramp down of the lattice depth. Because the gap of the amplitude mode is small in the crossover region [41], a relatively fast ramp-down across the crossover excites the amplitude mode. In contrast, such an oscillation is not observed in the experiment likely because of the combined effect of quantum and thermal fluctuations, and the spatial inhomogeneity due to the trap potential. Specifically for the inhomogeneity, the frequency of the amplitude mode significantly depends on the chemical potential, which varies in space in the presence of a trap potential, and this leads to the dephasing of the oscillation.

\section{Unexpectedly large atom loss for strong intensity of PA laser}
In this section, we discuss the possible origin of the unexpectedly large atom loss rate $\kappa$  observed for much higher intensity of PA laser corresponding to $\gamma > 5$.
This additional atom loss which is not taken into account in the present theory prevents the suppression of two-body loss rate from clear observation, and is also observed in the ramp-down dynamics in the deep lattice region.
For $V_0 = 20~E_R$, the measured loss rate is about 30 Hz, while the loss rate expected from the theory is about 3 Hz.
We confirm that this loss is not attributed to the photon scattering: the photon scattering rate we measure is about 3 Hz for intensity $I \sim 30$~W/cm$^2$.

Our calculation shows that Raman-assisted tunneling~\cite{aidelsburger-11} 
due to the PA laser can explain the observed additional loss.
Our high intensity PA laser not only induces the molecular formation but also the coupling between ${}^1 S_0$ and ${}^3 P_1$ atomic states with the detuning of 3.7 GHz, which results in the enhanced atomic tunneling by Raman process even though the overlap of the Wannier functions between the nearest neighboring site is small.
From our calculation, for $\gamma \sim 5$ in the lattice depth of $V_0 = 20~E_R$, this enhancement amounts to the two-body loss rate $\kappa \sim 50$~Hz, which is consistent with observation.
To avoid the effect of this additional loss, we restrict the region of the dissipation strength under $\gamma \sim 5$ in this work.

\clearpage

\begin{figure}[tbp]
\includegraphics[scale=0.8]{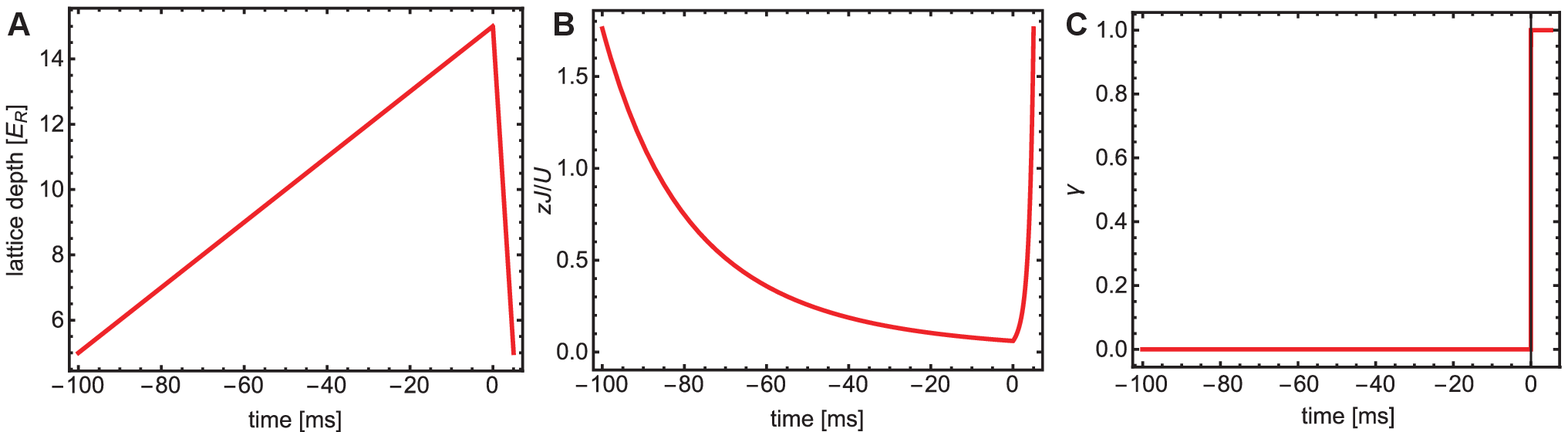}
  \caption{
\textbf{Time sequence for the dynamical melting of the Mott insulating state with unit filling.} 
\textbf{(A)}, The lattice depth, which is expressed in Eq.~(\ref{eq:Vsequence2}). \textbf{(B)}, $zJ/U$. \textbf{(C)}, $\gamma =\Gamma_{\rm PA}/U$ in the case that $\gamma(t>0)=1$.
  }
\label{fig:sequence2}
\end{figure} 

\begin{figure}[tbp]
\includegraphics[scale=0.62]{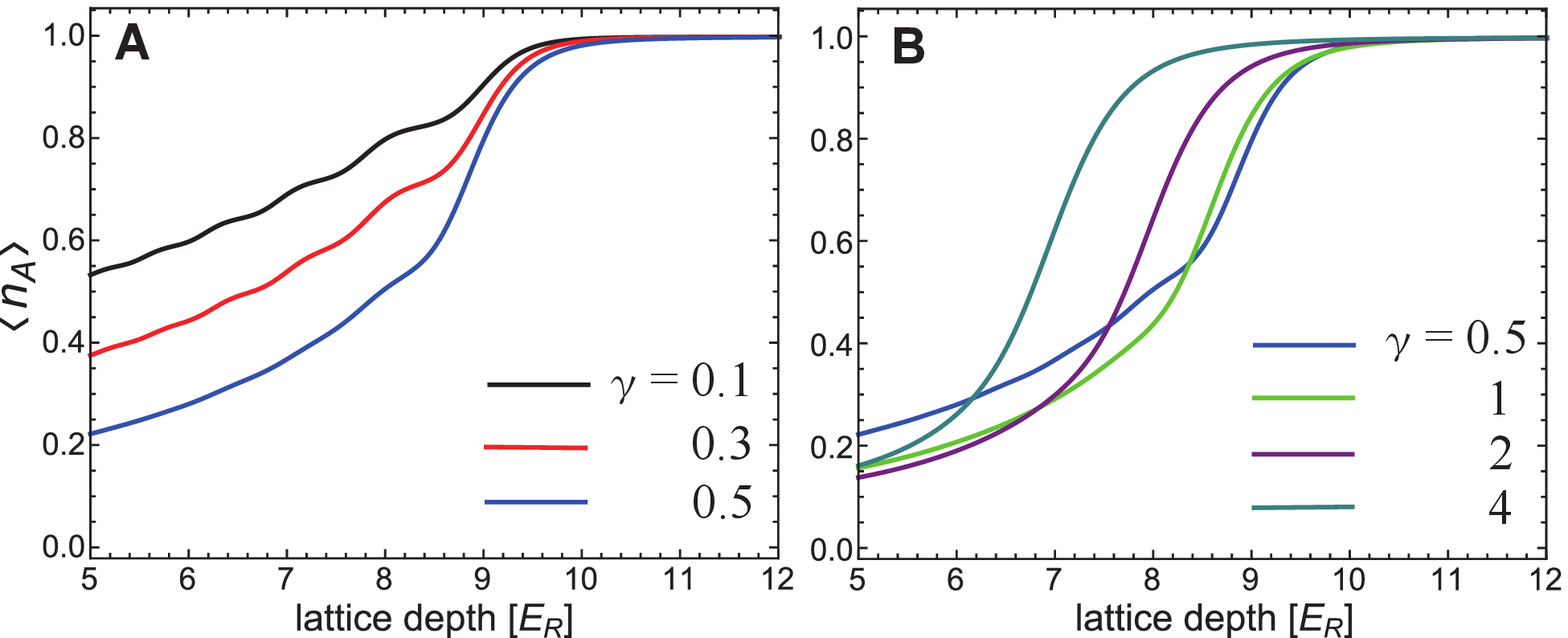}
  \caption{
\textbf{Atom density $\langle \hat{n}_{\rm A} \rangle$ as a function of the instantaneous value of the lattice depth $V_0/E_{\rm R}$.} 
We take the time region $t_1 \leq t < t_2$. 
Each curve represent a cross-section view of Fig.~2 (A) of the main text.
  }
\label{fig:nOfV}
\end{figure} 

\begin{figure}[tbp]
\includegraphics[scale=0.7]{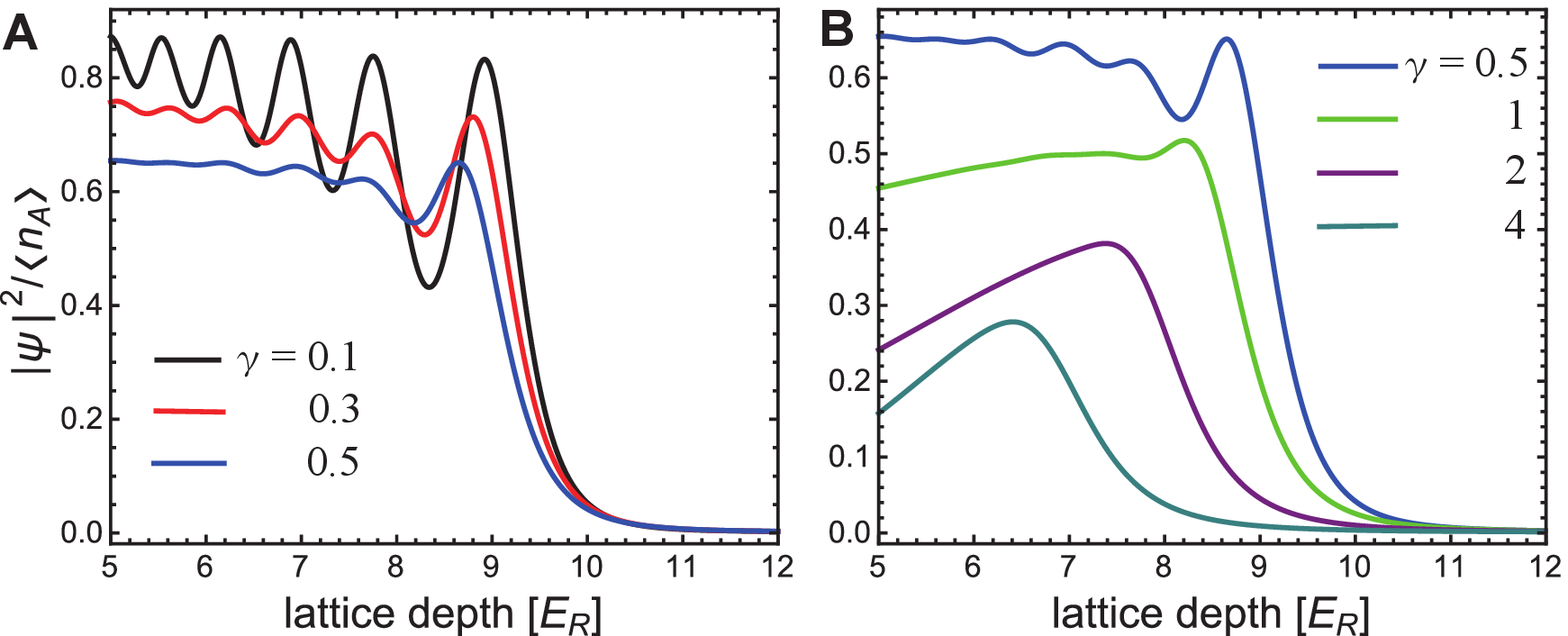}
  \caption{
\textbf{Condensate fraction $|\psi|^2/\langle \hat{n}_{\rm A}\rangle$ as a function of the instantaneous value of the lattice depth $V_0/E_{\rm R}$.} 
We take the time region $t_1 \leq t < t_2$. 
Each curve represent a cross-section view of Fig.~2 (A) of the main text.
  }
\label{fig:ncOfV}
\end{figure} 
%
%

\end{document}